\documentclass[prd,amsmath,amssymb,showpacs,floatfix,
superscriptaddress,nofootinbib,preprintnumbers]{revtex4-1}
\usepackage[latin1]{inputenc}
\usepackage{amsmath}
\usepackage{hyperref}
\usepackage{amsfonts}
\usepackage{amssymb}
\usepackage{graphicx}
\usepackage{appendix}
\usepackage{dcolumn}
\usepackage{caption}
\usepackage{subfigure}
\usepackage{float}
\usepackage{bm}
\usepackage{epsfig}

\def\eq#1{{Eq.~(\ref{#1})}}
\newcommand{\nn}{\nonumber \\}

\begin{document}

%opening
\title{Comparison of CMB lensing efficiency of gravitational waves and large scale structure}
\author{Hamsa Padmanabhan\footnote{email: hamsa@iucaa.ernet.in}, Aditya Rotti\footnote{email: aditya@iucaa.ernet.in} and Tarun Souradeep\footnote{email: tarun@iucaa.ernet.in} \\
IUCAA, Pune University Campus, Pune 411007, India
}
\begin{abstract}
We provide a detailed treatment and comparison of the weak lensing effects due to large-scale structure (LSS), or scalar density perturbations and those due to gravitational waves(GW) or tensor perturbations, on the temperature and polarization power spectra of the Cosmic Microwave Background (CMB). We carry out the analysis both in real space by using the correlation function method, as well as in the spherical harmonic space. We find an intriguing similarity between the lensing kernels associated with LSS lensing and GW lensing. It is found that the lensing kernels only differ in relative negative signs and their form is very reminiscent of even and odd parity bipolar spherical harmonic coefficients. Through a numerical study of these lensing kernels, we establish that lensing due to GW is more efficient at distorting the CMB spectra as compared to LSS lensing, particularly for the polarization power spectra. Finally we argue that the CMB B-mode power spectra measurements can be used to place interesting constraints on GW energy densities.

%Our numerical results establish the equivalence of the power spectra obtained by these two independent approaches. We are thus able to describe completely the weak lensing effects due to a stochastic gravitational wave background on the CMB.
\end{abstract}
\maketitle

\section{Introduction}

The weak gravitational lensing of the Cosmic Microwave Background (CMB) radiation due to large-scale structure (LSS) in the universe has been well studied in the literature. Weak gravitational lensing affects both the temperature anisotropy power spectrum as well as the polarization power spectra of the CMB.  In this article, we provide a quantitative treatment of the lensing effects due to gravitational waves (GW) that are the tensor metric perturbations, and make a comparative study of the lensing effects due to LSS and GW. This analysis is carried out in both, real space (by following the correlation function method as outlined in \cite{challinor} and extending the approach to include lensing due to GW's) as well as in harmonic space. We find that there exist interesting similarities in the structure of the lensing kernels associated with LSS  and GW, for the case of temperature as well as for the $E$ and $B$ modes of CMB polarization. We correct errors in the lensing kernels in existing literature \cite{cooray1}, which are used to derive modifications to the polarization power spectra due to lensing by GW. The robustness of these new results is established by checking that the temperature and polarization power spectra derived using the full-sky harmonic space analysis match those obtained by the independent real-space correlation function analysis. We are thus able to completely address the case of lensing by GW and provide an illuminating comparison that brings forth interesting similarity in the form of the expression with the complementary case of lensing due to LSS.

The plan of the paper is as follows: In Section \ref{sec:introlensing}, we start with a brief introduction to the theory of weak lensing and we introduce necessary terminology which is extensively used in the rest of the article. In Section \ref{sec:similarity}, we begin by briefly describing the outline of evaluating the lensing modifications to the CMB angular power spectra, following which, we discuss the curious similarities that exist in the structures of the lensing kernels arising from lensing due to LSS and GW. Since we correct the full-sky harmonic-space lensing kernels, we provide explicit details on the calculations in Appendixes \ref{app:polharspace2} and \ref{app:polharspace1} . For the full sky real space lensing kernels, we only outline the difference in calculation for LSS and GW lensing in the main body of the paper. For completeness, however, we present the complete derivation of the real space lensing kernels in the flat sky approximation, in the Appendix \ref{correlation-function-method-full-sky} .  In Section \ref{sec:results}, we discuss the numerical evaluations of the lensing kernels. This numerical study allows us to make a comparative study of the lensing kernels arising due to lensing due to LSS and GW. We summarize our findings and discuss implications of the results in the final section.

\section{Lensing by scalar and tensor perturbations}
\label{sec:introlensing}

The CMB photons arrive at us starting from the surface of last scattering, where the optical depth for these photons reaches a null. The CMB photons then transverse through a perturbed metric before reaching redshift $z=0$, where they are measured. Owing to these perturbations in the cosmological metric, the trajectories of the CMB photons deviate mildly from the ones that would have been in an unperturbed metric. The perturbations in the cosmological metric can be decomposed into two independently evolving parts, namely, scalar density perturbations which are associated with the LSS in the universe and the tensor perturbations which are associated with GWs in the universe. Both these components are responsible for perturbing the photon trajectories, however they differ in the nature of deflections that they introduce. The resultant phenomena is termed as lensing and has been shown to leave measurable imprint on the two point and three point correlation function of the CMB temperature anisotropy and polarization fields.

Consequently, weak lensing of the CMB photons results in a remapping of the CMB temperature anisotropies $T$ and the CMB polarization field ${}_\pm \tilde{X}$,
\begin{eqnarray}
\tilde{T}(\mathbf{\hat{n}} )=T(\mathbf{\hat{n}} +\bm{\alpha}) \,,\label{remap} \\
{}_\pm \tilde{X}(\mathbf{\hat{n}}) = {}_\pm X(\mathbf{\hat{n}} +\bm{\alpha}) \,.\label{remappol}
\end{eqnarray}
where $\bm{\alpha}$ is a vector field on the celestial sphere which denotes the deflection in the direction of photon arrival. The object describing the CMB polarization is defined as ${}_\pm X(\mathbf{\hat{n}}) = Q(\mathbf{\hat{n}}) \pm iU(\mathbf{\hat{n}})$, where $Q(\mathbf{\hat{n}})$ and $U(\mathbf{\hat{n}})$ are the standard Stokes parameters.

Since $\bm{\alpha}$ is a vector field, it can be decomposed into a gradient and a curl part \cite{stebbins},
\begin{equation}
\bm{\alpha}  = \nabla\psi(\mathbf{\hat{n}}) + \nabla\times\Omega(\mathbf{\hat{n}})  \,, %\sum_{lm} \psi_{lm} Y_{lm} (\mathbf{\hat{n}}) \,.
\label{psiexp}
\end{equation} 
where $\psi$ and $\Omega$ are scalar and pseudo-scalar fields defined on the sphere respectively. Note that, here the gradient and the curl operators are defined on the surface of the sphere which is orthogonal to the line of sight direction.

It has been shown that  at first order in perturbations, scalar metric perturbations only contribute to the gradient type modes, whereas tensor metric perturbations can induce both gradient and curl type modes in the deflection vector field \cite{cooray}. At second order in the perturbation, it is known that scalar metric perturbations do induce curl type deflections in photon trajectories. In this article, we limit the analysis to only linear order effects. Even though GWs induce both gradient and curl type deflections in photon trajectories at leading order in the perturbation, in this article we discuss the curl type deflections introduced due to GW. % (ignoring the gradient type deflection introduced due to GW). 

To evaluate the lensing modification to the CMB angular power spectrum requires prior knowledge about the two point statistic of the photon deflection vector field. Under the assumption of isotropy, the two point statistic for the photon deflections is completely characterized by specifying the angular power spectrum for the scalar field $\psi(\mathbf{\hat{n}})$ and the pseudo-scalar field $\Omega(\mathbf{\hat{n}})$. Given the initial power spectrum of the metric perturbations, one can arrive at the statistics of the photon deflection by solving the photon geodesic equation in the perturbed metric.
\begin{figure}
 \begin{center}
\includegraphics[height=8cm,width=10cm,angle=0]{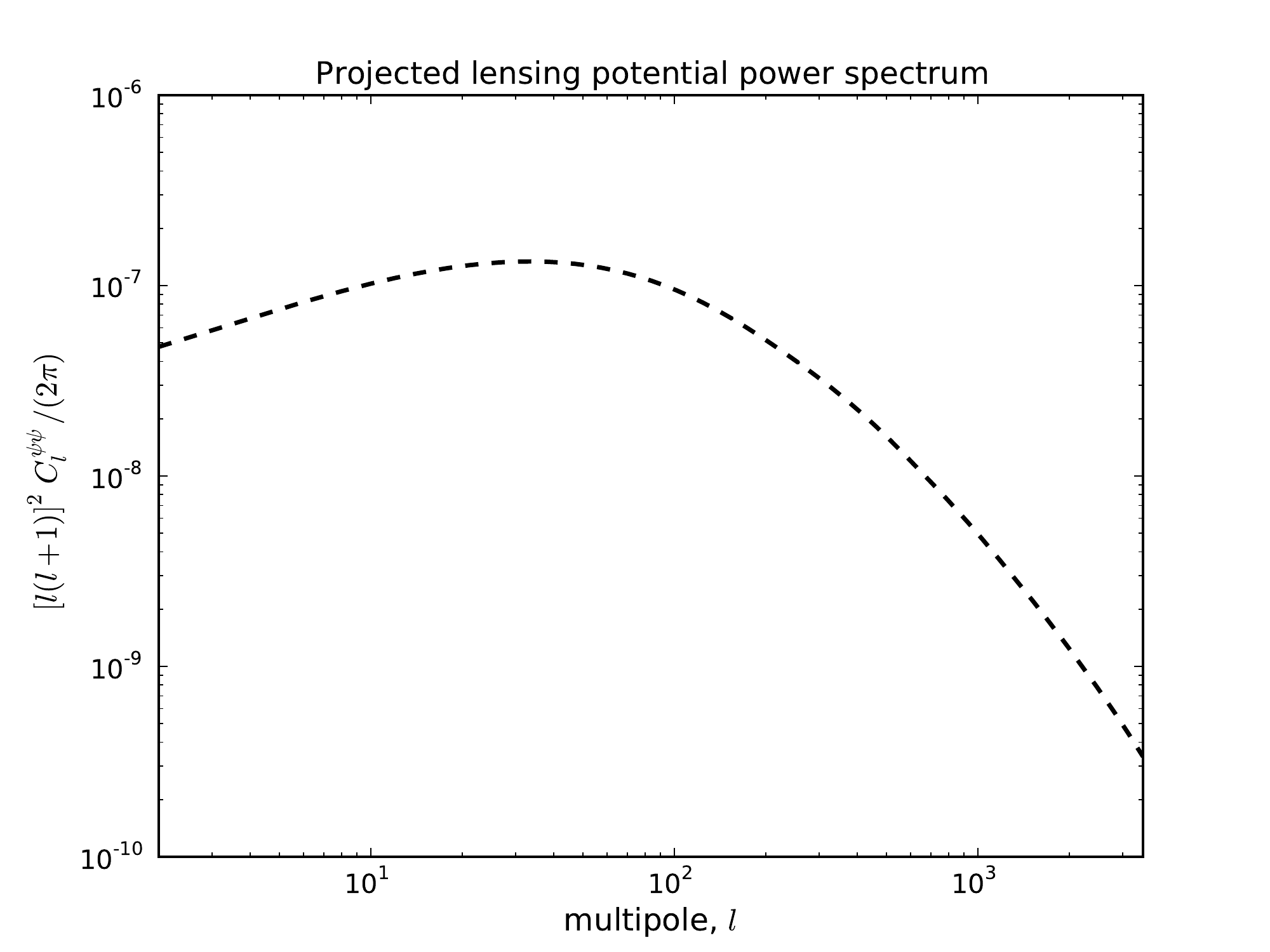} \end{center}
\captionsetup{singlelinecheck=off,justification=raggedright}
\caption{The lensing potential power spectra for the case of lensing by scalar perturbations that give rise to large scale structure in the universe. Figure drawn with numerical computations using CAMB\cite{camb}.}
\label{fig:lenspot1}
\end{figure} 
Specifically, scalar density perturbations associated with the intervening large-scale structure are characterized in terms of the Weyl potential, $\Psi_W$.
It can be shown, by solving the geodesic equation for the photon trajectory, that the scalar field $\psi$ introduced in \eq{psiexp}, is related to the Weyl potential through the following equation,
 \begin{equation}
\psi(\mathbf{\hat{n}}) = -2 \int_{0}^{r_{ls}} dr \frac{d_A(r_{ls}, r)}{d_A(r,0) d_A(r_{ls}, 0)}  \Psi_W(r, \mathbf{\hat{n}}) \,,
\label{psidef}
\end{equation} 
where the $d_A$'s denote the angular diameter distances, $r$ denotes the radial coordinate distance of the gravitational potential and $r_{ls}$ denotes the radial co-ordinate distance to the last scattering surface. Note that the scalar field $\psi$ is defined by taking a weighted sum of the gravitational potentials encountered by the CMB photon along its trajectory, hence allowing for its interpretation as an effective lens encountered by the CMB photons. In lensing literature the scalar field $\psi$ is better known as the projected lensing potential. Here, we reiterate that GW also contribute to $\psi$ but we ignore that contribution in the study presented in this article.

Likewise, the tensor metric perturbations are characterized by the transverse-traceless symmetric tensor $\mathcal{\textbf{H}}$, representing GWs.
Analogous to the manner in which the projected lensing potential $\psi$ was defined in terms of the perturbed metric variable $\Psi$,  it is possible to define \cite{dodelson} a projected lensing potential $\Omega (\mathbf{\hat{n}})$ in terms of the perturbed metric variable $\mathcal{\textbf{H}}$,
\begin{equation}
\label{omegadef}
 \Omega (\mathbf{\hat{n}}) = -\frac{1}{2} \int_{0}^{r_{ls}} dr \left[\mathbf{\hat{n}} \cdot (\nabla \times \mathcal{\textbf{H}}) \cdot \mathbf{\hat{n}} \right]_{(r, \mathbf{\hat{n}} r) } \,.
\end{equation} 
These cosmological stochastic GW could have many possible sources which include pre-recombination mechanisms such as inflation, phase transitions, ekpyrotic models, etc ... and post-recombination mechanisms such as cluster merges, cosmic strings etc ... 

It can be shown that the angular power spectra characterizing the gradient and curl component of deflections are related to the power spectrum of the metric perturbation variables $P_{\Psi}$ and $P_{\mathbf{H}}$, through the following equation,
\begin{equation}
\label{lens-spectrum}
 C_{l}^{x} = \frac{2}{\pi} \int dk \ k^2 P_{X}(k) \left| \mathcal{T}_l^{X}(k)\right|^2 \, ,
\end{equation}
where $x\equiv \lbrace \psi,\Omega \rbrace$ with corresponding $X\equiv \lbrace \Psi,\mathbf{H} \rbrace$, $k$ denotes the wavenumber and  $\mathcal{T}_l^{X}(k)$ accounts for the cosmic evolution of the respective perturbations and their projections onto the sphere. 

In the current cosmology model, there is a fairly good understanding of the model of formation of large scale structure, consequently, the expected $C_l^{\psi}$'s can be predicted. A plot of the lensing power spectrum $C_{l}^{\psi}$ is shown in Figure~\ref{fig:lenspot1} for a fiducial cosmology. Also the power spectrum predicted for the scalar density perturbations has been well constrained from the measurements of the matter power spectrum and more recently by PLANCK by directly measuring the cosmic lens \cite{planck-lens}.

\section{Similarity between scalar and tensor lensing kernels}
\label{sec:similarity}
In this section we study and clearly derive the curious similarity in the lensing kernels arising due to lensing by LSS and GW.

\subsection{Spherical harmonic space method}
This full-sky spherical harmonic space approach to arrive at the lensed CMB spectra has been extensively studied in Refs. \cite{waynehu, cooray1}. We briefly outline the calculations to arrive at the lensed CMB angular power spectra which characterize the CMB temperature anisotropies and polarization using the spherical harmonic space approach. We extend the harmonic space approach to derive the CMB lensed spectra to the case of lensing by GW, in process correcting lensing kernels previously derived. Finally we discuss an underlying similarity in the structure of the  harmonic space lensing kernels arising due to these two cases.

\subsubsection{Lensed temperature anisotropy power spectra}
The CMB temperature anisotropies form a scalar random field on the sphere and can be expanded in terms of spherical harmonic functions,
\begin{equation}
T(\mathbf{\hat{n}})
= \sum_{lm} a_{lm} Y_{lm}(\mathbf{\hat{n}}) \, .   
\label{tempylm}                               
\end{equation}
Under the assumption of Gaussianity and isotropy of these fluctuations, these temperature anisotropies are completely characterized by the angular power spectrum defined as, $ C_l^{TT} = \langle |a_{lm} |^2 \rangle$.

Weak lensing of the CMB photons results in this field getting remapped as described in \eq{remap}. As a result of this remapping, the statistics of the fluctuations of the CMB temperature anisotropies are altered. To arrive at an expression for the lensed CMB temperature anisotropy power spectrum, we begin by expressing the lensed temperature anisotropies as a function of the unlensed temperature anisotropies and the deflection field $\bm{\alpha}$.   This is achieved by expanding the lensed temperature anisotropies in a Taylor series (\eq{taylorexptemp}) in the variable $\bm{\alpha}$, as follows,
\begin{equation}
 \widetilde{T}(\mathbf{\hat{n}}) = T(\mathbf{\hat{n}}+\bm{\alpha}) 
 =T(\mathbf{\hat{n}})+\nabla^a T\cdot
\alpha_a+\frac{1}{2}\nabla^b\nabla^a T\cdot\alpha_a\alpha_b + ... \, 
\label{taylorexptemp}
\end{equation}
The next step is to transform the derived expression to its spherical harmonic space equivalent. On manipulating the resulting harmonic space expression, it can be shown that the lensed temperature angular power spectrum $\widetilde{C}_l^{TT}$, is given by the following expression \cite{waynehu,cooray1},
\begin{equation}
 \widetilde{C}_l^{TT} = C_l^{TT}- l(l+1) R C_l^{TT}+ \sum_{l_1 l_2} \frac{C_{l_1}^{TT}}{2 l + 1} [C_{l_2}^{\psi} |F^{\psi}_{l l_1 l_2}|^2 + C_{l_2}^{\Omega} |F^{\Omega}_{l l_1 l_2}|^2]
 \label{lensedclharmonic}
\end{equation}
where $C_{l}^{\psi}$ and $C_{l}^{\Omega}$ are the lensing potential power spectra introduced previously, in \eq{lens-spectrum}. $C_{l_1}^{TT}$ is the unlensed CMB power spectrum and $R$ denotes the root mean square deflection power and is related to the deflection power spectra through the following expression,
\begin{equation}
 R = \sum_{l} \frac{l (l +1) (2 l + 1)}{8 \pi} [C_{l}^{\psi} + C_{l}^{\Omega}] \,.
 \label{rmsdefpowdisc}
\end{equation} 
The functions $F^\psi$ and $F^\Omega$  in \eq{lensedclharmonic} are the lensing kernels that describe the effect of the lensing potential on the unlensed CMB anisotropy angular power spectrum, and hence these functions act as \textit{kernels}. The explicit expression for the kernel in the case of lensing due to LSS is given by \cite{waynehu},
\begin{equation}
F_{ll_1l_2}^\psi=\frac{1}{2}\left[l_1(l_1+1)+l_2(l_2+1)-l(l+1)\right]
  \sqrt{\frac{(2l+1)(2l_1+1)(2l_2+1)}{4\pi}} 
  \begin{pmatrix}
l&l_1&l_2\\
0&0&0
\end{pmatrix} \,.
\label{fplus} 
\end{equation} 
The expressions in brackets are the Wigner$-3j$ symbols. Using the standard Wigner-3j identity, 
\begin{equation}
[l_2(l_2 + 1) + l_1 (l_1 + 1) - l(l+1)] \begin{pmatrix}
l&l_1&l_2\\
0&0&0
\end{pmatrix} 
=-\sqrt{l_2(l_2 + 1)l_1 (l_1 + 1)} [1 + (-1)^{l + l_1 + l_2}] \begin{pmatrix}
l&l_1&l_2\\
0&-1&1
\end{pmatrix} \,,
\end{equation} 
the \eq{fplus} can be recast in the following form,
\begin{equation}
F_{ll_1l_2}^\psi= -\frac{1}{2}\sqrt{l_1(l_1+1)l_2(l_2+1)}
\begin{pmatrix}l&l_1&l_2\\ 0&-1&1\end{pmatrix} 
 \sqrt
{\frac{(2l+1)(2l_1+1)(2l_2+1)}{4\pi}} 
 \left[1+(-1)^{l+l_1+l_2}\right] \,.
\label{fplusmodified}
\end{equation}
and the expression for the kernel in the case of lensing due to GW is given by \cite{cooray} ,
\begin{equation}
F_{ll_1l_2}^\Omega=\frac{1}{2}\sqrt{l_1(l_1+1)l_2(l_2+1)}
\begin{pmatrix}l&l_1&l_2\\ 0&-1&1\end{pmatrix} 
  \sqrt{\frac{(2l+1)(2l_1+1)(2l_2+1)}{4\pi}} 
 \left[1-(-1)^{l+l_1+l_2}\right] \,.
\label{fcross}
\end{equation}

On inspecting \eq{fplusmodified} and \eq{fcross}, it can be seen that the lensing kernel for the case of GW lensing is very similar in structure to the  corresponding kernel for LSS lensing and differs from the latter only by a relative minus sign between the two terms, i.e. it contains the factor $[1-(-1)^{l+l_1+l_2}]$ as opposed to $[1+(-1)^{l+l_1+l_2}]$ for the corresponding kernel for the case of LSS lensing.
(The overall negative sign in front of \eq{fplusmodified}  is not crucial, since only the squares of the $F$-functions appear in the lensing kernel.)  Also, note that the kernel $F_{ll_1l_2}^\psi$ is non-vanishing only for the case of even $l+l_1+l_2$, while that for $F_{ll_1l_2}^\Omega$ is non-vanishing for the exactly complementary case of odd $l+l_1+l_2$. It is likely that this behaviour is closely linked to the parity properties for these two cases, as indicated by the lensing studies in the Bipolar Spherical Harmonic representation\cite{BookMKTS2012} and in \cite{namikawa}. 

\subsubsection{Lensed polarization power spectra}

Just as in the case of lensing of CMB temperature anisotropy, we analyze the lensing modification to the angular power spectra which characterize the CMB polarization field. The object describing CMB polarization ${}_\pm X(\mathbf{\hat{n}})$ introduced in \eq{remappol} forms a spin-2 field on the sphere.
Analogous to spherical harmonic decomposition of the CMB temperature anisotropies, ${}_\pm X(\mathbf{\hat{n}})$ can be expanded in terms of spin-weighted spherical harmonics on the sky as follows,
\begin{equation}
 {}_{\pm} X(\mathbf{\hat{n}}) = \sum_{lm} {}_{\pm} a_{lm}^X {}_{\pm 2} Y_{lm}(\mathbf{\hat{n}}) \,,
\end{equation} 
where $_{\pm 2} Y_{lm}(\mathbf{\hat{n}})$ are the spin-2 spherical harmonics \cite{goldberg}. The Stokes parameters and hence the polarization field  ${}_\pm X(\mathbf{\hat{n}})$ is a coordinate dependent quantity. Since it is difficult to work with coordinate dependent quantities, it is common to introduce the $E$-field and the $B$-field which are scalar fields and provide an equivalent description of CMB polarization. These fields are constructed such that they are coordinate independent fields and they are related to the polarization field  ${}_\pm X(\mathbf{\hat{n}})$  through the following spherical harmonic space expression,
\begin{equation}
_{\pm} a_{lm}^X = a_{lm}^E \pm i a_{lm}^B
\end{equation} 
where $a_{lm}^E$ and $a_{lm}^B$ are the spherical harmonic coefficients of the field $E$ and $B$ respectively. 
The power spectra of the polarization modes are defined, analogously to the temperature, as $C_l^{EE} = \langle |a_{lm}^E |^2 \rangle, C_l^{BB} = \langle |a_{lm}^B |^2 \rangle$.

To arrive at the lensed CMB polarization angular power spectra, one needs to expand the lensed polarization field ${}_\pm \tilde{X}(\mathbf{\hat{n}})$ in a Taylor series and then transform the equation into its harmonic space equivalent. This procedure leads to the following expression,

\begin{eqnarray}
\label{lensed_pol_alm}
&& {}_\pm\tilde{a}_{lm}^X = {}_\pm
a_{lm}^X - \sum_{l_1m_1l_2m_2} \Big[{}_\pm a_{l_2m_2}^X \Big(
{}_{\pm 2}I^\psi_{lml_1m_1l_2m_2}\psi_{l_1m_1}
 + {}_{\pm 2} I^\Omega_{lml_1m_1l_2m_2}\Omega_{l_1m_1}\Big)\Big]
 \nonumber\\ 
 &+&\frac{1}{2}\sum_{l_1m_1l_2m_2l_3m_3} \Big[{}_\pm
a_{l_2m_2}^X \Big({}_{\pm
2}J^\psi_{lml_1m_1l_2m_2l_3m_3}\psi_{l_1m_1}\psi^{
*}_{l_3m_3}  \nonumber\\
&+& {}_{\pm
2}J^\Omega_{lml_1m_1l_2m_2l_3m_3}\Omega_{l_1m_1}\Omega^{
*}_{l_3m_3}\Big)\Big] \, ,
\end{eqnarray}
where,
\begin{eqnarray}
\label{pol_integrals}
 _{\pm 2}I_{lml_1m_1l_2m_2}^\psi &=&\int d\mathbf{\hat{n}} \ _{\pm 2}Y_{lm}^* \nabla_a Y_{l_1m_1} \nabla^a _{\pm 2} Y_{l_2m_2}\nonumber\\
_{\pm 2} I_{lml_1m_1l_2m_2}^\Omega &=&\int  d\mathbf{\hat{n}}  \ _{\pm 2} Y_{lm}^* \nabla^a _{\pm 2}Y_{l_2m_2} \nabla_b Y_{l_1m_1}\epsilon^b{}_a \nonumber\\
_{\pm 2}J_{lml_1m_1l_2m_2l_3m_3}^\psi &=& \int d\mathbf{\hat{n}} \
_{\pm 2}Y_{lm}^* \nabla^a \nabla^b _{\pm 2}Y_{l_2m_2} \nabla_a Y_{l_1m_1} \nabla_b  Y_{l_3m_3}^* \nonumber \\
_{\pm 2} J_{lml_1m_1l_2m_2l_3m_3}^\Omega &=& \int d\mathbf{\hat{n}} \
_{\pm 2}Y_{lm}^*\nabla^a \nabla^b _{\pm 2}Y_{l_2m_2} \nabla_c Y_{l_1m_1} \nabla_d Y_{l_3m_3}^*\epsilon^c{}_a\epsilon^d{}_b
\, . 
\end{eqnarray}

To arrive at the expressions for the lensed polarization power spectra and the polarization-temperature cross power spectra one needs to evaluate the corresponding covariance matrices starting from \eq{lensed_pol_alm} and using results of the explicit evaluations of the integrals in \eq{pol_integrals}. The details of the derivation are provided in Appendix \ref{app:polharspace1}.

Finally, it can be shown that the lensed $E$ and $B$ mode power spectra are given by the following expressions,
\begin{eqnarray} \label{correctedcls}
\tilde{C}_l^{EE}&=&C_{l}^{EE}-(l^2+l-4)RC_l^{EE}+\sum_{l_1 l_2 X} \frac{\left[(C_{l_1}^{EE}+C_{l_1}^{BB})+(-1)^{L^{X}}(C_{l_1}^{EE}-C_{l_1}^{BB}) \right]}{2(2l+1)} C_{l_2}^{X}|{}_2F_{l l_1 l_2}^{X}|^2 \,, \label{Corr_ker_1}
\end{eqnarray}
\begin{eqnarray}
\tilde{C}_l^{BB}&=&C_{l}^{BB}-(l^2+l-4)RC_l^{BB}+\sum_{l_1 l_2 X} \frac{\left[(C_{l_1}^{EE}+C_{l_1}^{BB})-(-1)^{L^{X}}(C_{l_1}^{EE}-C_{l_1}^{BB}) \right]}{2(2l+1)} C_{l_2}^{X}|{}_2F_{l l_1 l_2}^{X}|^2  \label{Corr_ker_2}\,,\\
\tilde{C}_l^{TE}&=&C_{l}^{TE}-(l^2+l-2)RC_l^{TE}+\sum_{l_1 l_2}\frac{C_{l_1}^{TE}}{2l+1} \left[C_{l_2}^{\psi}|F_{l l_1 l_2}^{\psi}||{}_2F_{l l_1 l_2}^{\psi}| - C_{l_2}^{\Omega}|F_{l l_1 l_2}^{\Omega}||{}_2F_{l l_1 l_2}^{\Omega}|\right] \label{Corr_ker_3}\,.
\end{eqnarray}
\noindent where, $X  \equiv \lbrace \psi , \Omega \rbrace$, $L^\psi=l+l_1+l_2$,  $L^\Omega=l+l_1+l_2+1$, $R$ is the root mean square deflection power defined in \eq{rmsdefpowdisc}, the functions $F^\psi_{ll_1l_2}$ \& $F^\Omega_{ll_1l_2}$ are defined in \eq{fplusmodified} \& \eq{fcross} respectively and $_{2}F^\psi_{ll_1l_2}$ and $_{2}F^\Omega_{ll_1l_2}$ are given by the following expressions,
\begin{equation}
\label{pol_psi_old}
{}_{2}F_{ll_1l_2}^\psi=\frac{1}{2}[l_1(l_1+1)+l_2(l_2+1)-l(l+1)] 
\sqrt{\frac{(2l+1)(2l_1+1)(2l_2+1)}{4\pi}}\begin{pmatrix}l&l_1&l_2\\2&0&-2\end{pmatrix}
\end{equation} 
and
\begin{eqnarray}
\label{2fomega}
{}_{2}F^\Omega_{ll_1l_2}&=&\sqrt{\frac{l_1(l_1+1)(2l+1)(2l_1+1)(2l_2+1)}{8\pi}}\\
&&\hskip-3em \times\ \Bigg[
\sqrt{\frac{(l_2+2)(l_2-1)}{2}}\begin{pmatrix}l&l_1&l_2\\2&-1&-1\end{pmatrix}
 -\sqrt{\frac{(l_2-2)(l_2+3)}{2}} \begin{pmatrix}l&l_1&l_2\\2&1&-3\end{pmatrix}\Bigg]\nonumber  
 \end{eqnarray}

On evaluating the integrals given in \eq{pol_integrals}, we find the following mathematical identities to hold true,  $_{\mp 2} I^{\psi}_{lml_1m_1l_2m_2} = (-1)^{L^{\psi}} _{\pm 2} I^{\psi}_{lml_1m_1l_2m_2}$ and $_{\mp 2} I^{\Omega}_{lml_1m_1l_2m_2} = (-1)^{L^{\Omega}}_{\pm 2} I^{\Omega}_{lml_1m_1l_2m_2}$,  where $L^{\Omega}$ is related to $L^{\psi}$ as $L^{\Omega}=L^{\psi}+1$. The derivation of these identities are outlined in Appendix \ref{app:polharspace2}. As a results of these identities, we find the lensing modifications to the CMB power spectra due to lensing by GW given in \eq{Corr_ker_1}, \eq{Corr_ker_2} \& \eq{Corr_ker_3} with $X \equiv \Omega $, which differ from the corresponding expressions given in existing literature \cite{cooray1}. Specifically, the term multiplying $C_{l_1}^{\Omega}({}_{2}F^\Omega_{ll_1l_2})^2$ on the right hand sides of the expressions for $\widetilde{C}_{l}^{EE}$ and $\widetilde{C}_{l}^{BB}$, is $[(C_{l_1}^{EE}+C_{l_1}^{BB})\mp(-1)^{L^\psi}(C_{l_1}^{EE}-C_{l_1}^{BB})]$, and not $[(C_{l_1}^{EE}+C_{l_1}^{BB})\pm(-1)^{L^\psi}(C_{l_1}^{EE}-C_{l_1}^{BB})]$, as has been previously mentioned in the literature. Similarly, the term multiplying $C_{l_1}^{\Omega}$ on the right hand side of the expression for $\widetilde{C}_{l}^{TE}$ is $-C_{l_1}^{\Omega}({}_{+2}F_{ll_1l_2}^\Omega)$, and not $+ C_{l_1}^{\Omega}({}_{+2}F_{ll_1l_2}^\Omega)$. These modified expressions match the corresponding ones obtained by a real-space correlation function (discussed in the following section) approach following Ref. \cite{challinor}, as shown in the results of Section \ref{sec:results}.

Finally we study the similarity in the lensing kernels which appear while evaluating the lensing modification to the CMB polarization power spectra. We show that the same behaviour as observed for the lensing kernels in case of the temperature power spectra is reproduced in this case as well. Specifically, the expression for the lensing kernel ${}_{2}F^\psi_{ll_1l_2}$ (given in \eq{pol_psi_old}) can be equivalently expressed in the following form,
\begin{eqnarray}
\label{2fpsiequiv}
{}_{2}F^\psi_{ll_1l_2}&=&\sqrt{\frac{l_1(l_1+1)(2l+1)(2l_1+1)(2l_2+1)}{8\pi}}\\
&&\hskip-3em \times\ \Bigg[
\sqrt{\frac{(l_2+2)(l_2-1)}{2}}\begin{pmatrix}l&l_1&l_2\\2&-1&-1\end{pmatrix}
 +\sqrt{\frac{(l_2-2)(l_2+3)}{2}} \begin{pmatrix}l&l_1&l_2\\2&1&-3\end{pmatrix}\Bigg] \,.\nonumber   
 \end{eqnarray} 
Inspecting the lensing kernels given in \eq{2fomega} \& \eq{2fpsiequiv}, it is observed that the two lensing kernels are again very similar in structure in this case as well, the only difference being a relative sign change between two terms. A derivation of the above expression is provided in Appendix \ref{app:polharspace2}.
%%%%%%%%%%%%%%%%%%%%%%%%%%%%%%%%%%%%%%%%%%%%%%%%%%%%%%%%%%
\subsection{Real space correlation function method : Full sky}
\label{corr-func-method}
The method described in the previous section is accurate only as long as the Taylor series expansion of the lensed CMB fields is valid. The Taylor series approximation begins to fail once the angular scales involved start becoming comparable to the magnitude of photon deflection induced due to lensing. This fact is demonstrated in a more quantitative fashion in Section~\ref{sec:results}.

This section discusses the real space correlation function method \cite{challinor} which allows for the evaluation of the lensed CMB spectra without making the Taylor series approximation to the lensed CMB fields. We outline the methodology to evaluate the LSS lensing modifications to the CMB spectra and then extend this method to the case of lensing by GW. For completeness, an exhaustive discussion of this method in the flat sky approximation is discussed in Appendix~\ref{correlation-function-method-full-sky}.

The unlensed temperature anisotropy correlation function  $\xi(\beta)$ is given by the following expression,
\begin{equation}
 \xi(\beta) = \langle T(\hat{\mathbf{n}}_1)T(\hat{\mathbf{n}}_2) \rangle
\end{equation} 
where $\hat{\mathbf{n}}_1$ and $\hat{\mathbf{n}}_2$ are the two directions and $\cos \beta = \hat{\mathbf{n}}_1 \cdot\hat{\mathbf{n}}_2$. 
The lensed temperature correlation function is defined as $\widetilde{\xi}(\beta) = \langle T(\hat{\mathbf{n}}'_1) T(\hat{\mathbf{n}}'_2) \rangle $, where $\hat{\mathbf{n}}'_1 = \hat{\mathbf{n}}_1 + \bm{\alpha}(\hat{\mathbf{n}}_1) \,,\hat{\mathbf{n}}'_2 = \hat{\mathbf{n}}_2 + \bm{\alpha}(\hat{\mathbf{n}}_2)\,$ and $\bm{\alpha}(\hat{\mathbf{n}}_1)$ and $\bm{\alpha}(\hat{\mathbf{n}}_2)$ are the lensing deflections at $\hat{\mathbf{n}}_1$ and $\hat{\mathbf{n}}_2$ respectively.
Similarly, we can define the lensed correlation functions for the $EE$, $BB$ and $TE$ polarization modes.

We follow Ref. \cite{challinor} in which a spin-1 deflection field  ${}_1 \alpha = \bm{\alpha} \cdot \sqrt{2} \mathbf{m}_-$ is introduced, where $\mathbf{m}_{\pm}$ are related to the standard basis functions on the sphere, $\mathbf{e}_{\theta}$ and $\mathbf{e}_{\phi}$ as follows,
\begin{equation}
 \mathbf{m}_{\pm} = \frac{1}{\sqrt{2}} (\mathbf{e}_{\theta} \mp  i\mathbf{e}_{\phi}) \, .
\end{equation}
The spin-1 deflection in the basis defined by the geodesic connecting $\hat{\mathbf{n}}_1$ and $\hat{\mathbf{n}}_2$ is denoted by an overbar, ${}_1\bar{\alpha}$.  We consider the correlation between the pairs of quantities ${}_1 \bar{\alpha}(\hat{\mathbf{n}}_1)$ and ${}_1\bar{\alpha}(\hat{\mathbf{n}}_2)$ as also ${}_1\bar{\alpha}^*(\hat{\mathbf{n}}_1)$ and ${}_1\bar{\alpha}(\hat{\mathbf{n}}_2)$. We define the functions,
\begin{equation}
C_{gl,2}(\beta) = \langle  {}_1 \bar{\alpha} (\hat{\mathbf{n}}_1)  {}_1 \bar{\alpha} (\hat{\mathbf{n}}_2) \rangle                                                                                              
\label{def-corr2}
 \end{equation} 
and
\begin{equation}
C_{gl}(\beta) = \langle  {}_1 \bar{\alpha^*} (\hat{\mathbf{n}}_1)  {}_1 \bar{\alpha} (\hat{\mathbf{n}}_2) \rangle \, .                                                                                              
\label{def-corr1}
\end{equation}

It can be shown that the lensed temperature and polarization correlation functions, can be expressed in terms of these functions, eventually leading to expressions for the lensed CMB angular power spectra. In this article we study the modifications to these quantities due to lensing by tensor perturbations \cite{challinor} . In Ref. \cite{challinor}, the expressions for the full-sky correlation functions for lensing due to large-scale structure, arising from  scalar perturbations, have been derived. We discuss the full-sky analysis for the case of lensing by tensor perturbations.
 The curl type deflection induced due to lensing by GW are expressed through the following expression,
\begin{equation}
 \bm{\alpha}_b = \epsilon^a{}_b \nabla_a \Omega     
\end{equation} 
where $\epsilon^a{}_b$ is the two dimensional totally antisymmetric Levi-Civita tensor. The spin-1 deflection filed can be written as follows, 
\begin{equation}
{}_1 \bar{\alpha} = \sqrt{2} \epsilon^a{}_b (\mathbf{m}_-)^b \nabla_a \Omega 
\end{equation} 
 Given the above expression, we can evaluate the two point correlation functions (\eq{def-corr2} \& \eq{def-corr1}) of the spin-1 deflection field induced due to lensing by gravitational waves (the subscript ``$T$'' stands for tensors) in terms of the power spectrum $C_l^{\Omega}$ of the field $\Omega$ defined in \eq{omegadef}.

The correlation functions of the spin-1 deflection field are given by the following expression,
\begin{eqnarray}
 -C_{gl,2}(\beta)_{T} \equiv \langle  {}_1 \bar{\alpha} (\hat{\mathbf{n}}_1)  {}_1 \bar{\alpha} (\hat{\mathbf{n}}_2) \rangle _{T} &=&  2 \sum_{l,m}   C_l^{\Omega} [\epsilon^a{}_b \nabla_a Y_{lm}(\hat{\mathbf{n}}_1) (\mathbf{m}_-)^b] [\epsilon^i{}_j \nabla_i \,  Y^*_{lm} (\hat{\mathbf{n}}_2)(\mathbf{m}_-)^j] \nonumber\\
&=& \sum_{l,m} C_l^{\Omega} l (l+1) \,{}_1 Y_{lm}(\hat{\mathbf{n}}_1) {}_{-1} Y^*_{lm}(\hat{\mathbf{n}}_2) \nonumber\\
&=& \sum_{l} \frac{(2l +1)}{4 \pi}l (l+1) C_l^{\Omega} d^{l}_{-11}(\beta)
\label{gwa2rfullsky}
\end{eqnarray}
where we have used the relation \cite{goldberg},
\begin{equation}
 \nabla_a Y_{lm} = \sqrt{\frac{l(l+1)}{2}} [{}_1 Y_{lm} (\mathbf{m}_+)_a - {}_{-1} Y_{lm} (\mathbf{m}_-)_a]
\end{equation} 
Similarly it can be shown that,
\begin{eqnarray}
 C_{gl}(\beta)_{T} \equiv \langle  {}_1 \bar{\alpha}^* (\hat{\mathbf{n}}_1)  {}_1 \bar{\alpha} (\hat{\mathbf{n}}_2) \rangle _{T} &=&  2 \sum_{l,m}   C_l^{\Omega} [\epsilon^a{}_b \nabla_a Y_{lm}(\hat{\mathbf{n}}_1) (\mathbf{m}_+)^b] [\epsilon^i{}_j \nabla_i   Y^*_{lm} (\hat{\mathbf{n}}_2)(\mathbf{m}_-)^j] \nonumber\\
&=& \sum_{l,m} C_l^{\Omega} l (l+1) {}_{-1} Y_{lm}(\hat{\mathbf{n}}_1) {}_{-1} Y^*_{lm}(\hat{\mathbf{n}}_2) \nonumber\\
&=& \sum_{l} \frac{(2l +1)}{4 \pi}l (l+1) C_l^{\Omega} d^{l}_{11}(\beta) \,.
\label{gwa0rfullsky}
\end{eqnarray} 

Following the notation in Ref. \cite{challinor}, the corresponding expressions for the case of lensing by  LSS or scalar metric perturbations are given by (subscript ``$S$'' stands for scalars),
\begin{equation}
-C_{gl,2}(\beta)_{S} \equiv \langle  {}_1 \bar{\alpha} (\hat{\mathbf{n}}_1) {}_1 \bar{\alpha} (\hat{\mathbf{n}}_2) \rangle _{S} = - \sum_{l} \frac{2 l + 1}{4 \pi} l(l+1) C_l^{\psi} d_{-11}^l
(\beta)
\label{scalara2rfullsky}
\end{equation} 
and
\begin{equation}
C_{gl}(\beta)_{S} \equiv \langle  {}_1 \bar{\alpha}^* (\hat{\mathbf{n}}_1) {}_1 \bar{\alpha} (\hat{\mathbf{n}}_2)\rangle _{S} =  \sum_{l} \frac{2 l + 1}{4 \pi} l(l+1) C_l^{\psi} d_{11}^l
(\beta)
\label{scalara0rfullsky}
\end{equation} 
\eq{gwa2rfullsky} and \eq{gwa0rfullsky} are the analogues of \eq{scalara2rfullsky} and \eq{scalara0rfullsky}, respectively, for the case of lensing by tensor perturbations. Hence, we can see that while the expression for $C_{gl}(\beta)$ has the same form for both the cases of lensing by scalar and tensor perturbations, the expression for $C_{gl, 2}(\beta)$ picks up a relative negative sign for lensing by tensor perturbations, as compared to that for scalar perturbations. This result ties in well with our previous findings in harmonic space where, again, we have seen that the gravitational wave contribution introduces a relative negative sign. These correlation functions derived for the deflection field can be used to arrive at the lensed CMB angular power spectra and the details of that can be found in the literature \cite{challinor}.
%%%%%%%%%%%%%%%%%%%%%%%%%%%%%%%%%%%%%%%%%%%%%%%%%%%%%%%%%%

%\input{./sections/section3}
\section{Numerical evaluation of the lensing kernels}
\label{sec:results}

In Section~\ref{sec:similarity}, we derive the lensing modifications to the CMB angular power spectra using the harmonic space approach and the real space approach, which are two completely independent methods. In this section of the article, we summarise the results of the numerical study of the lensing kernels. The real space lensing kernels due to lensing by gravitational waves have been derived for the first time in this article. Also, we have corrected the existing full sky harmonic space kernels associated with the curl type photon displacements. 

We perform a numerical study to establish the equivalence of these newly derived kernels in real and harmonic space for lensing due to GW and LSS. We also use this analysis to study how the curl type displacements and gradient type displacements of the photons differ in the manner in which they distort the CMB angular power spectra. 

For the numerical evaluation of the harmonic space kernels we have developed a Fortran code that computes the lensing kernels and gives the lensed CMB spectra as output. The results of this code have been calibrated against the lensed CMB spectra computed by \href{http://www.camb.info/}{CAMB} for the case of LSS lensing \footnote{Note that here we have changed the lensing switch in CAMB so that it uses the harmonic space kernels to evaluate the lensed CMB spectra.}. To evaluate the lensed CMB spectra using the real space lensing kernels, we have made minor edits to \href{http://www.camb.info/}{CAMB} (changing the sign of the function $A_2$ introduced in Section~\ref{corr-func-method}) to account for lensing due to GW.

In the lensing kernels, the only unknown is the power spectrum of the curl type displacements, $C_{l}^{\Omega\Omega}$. In principle, one can evaluate $C_{l}^{\Omega\Omega}$ given the power spectra of the tensor perturbations arising from all sources. Since the main motive is to make a comparative study of lensing by LSS and lensing  by GW, we choose the curl deflection power spectra $C_{l}^{\Omega\Omega}$ to be equal to the LSS projected lensing potential power spectra $C_{l}^{\psi\psi}$. Note that we consider independently the cases of lensing by LSS alone $(C_{l}^{\Omega\Omega}=0)$, vis-a-vis lensing by GW alone $(C_{l}^{\psi\psi}=0)$, to show the differences between these two cases. By doing so, we probe the differences between the lensing kernels which were discussed in Section~\ref{sec:similarity}. 
The results are summarised in the following subsections.

\subsection{Equivalence of harmonic space lensing kernels and real space lensing kernels}
\label{results-a}
The harmonic space approach relies on the Taylor series expansion of the lensed temperature and polarization fields. This Taylor series expansion is expected to become increasingly inaccurate as one approaches larger multipoles. This is due to the fact that at the corresponding angular scales, the amplitude of the transverse photon deflections become comparable to the angular size of the CMB sky under study.  The real space approach, however, is valid at all angular scales. As a consequence, the lensed spectra obtained from the harmonic space kernels deviate from those obtained from the real space calculations as one approaches larger multipoles. 
\begin{figure}[!h]
\begin{center}
%	\mbox{
    \subfigure[$TT$]{\label{fig2a}  \includegraphics[height=6cm,width=8cm,angle=0]{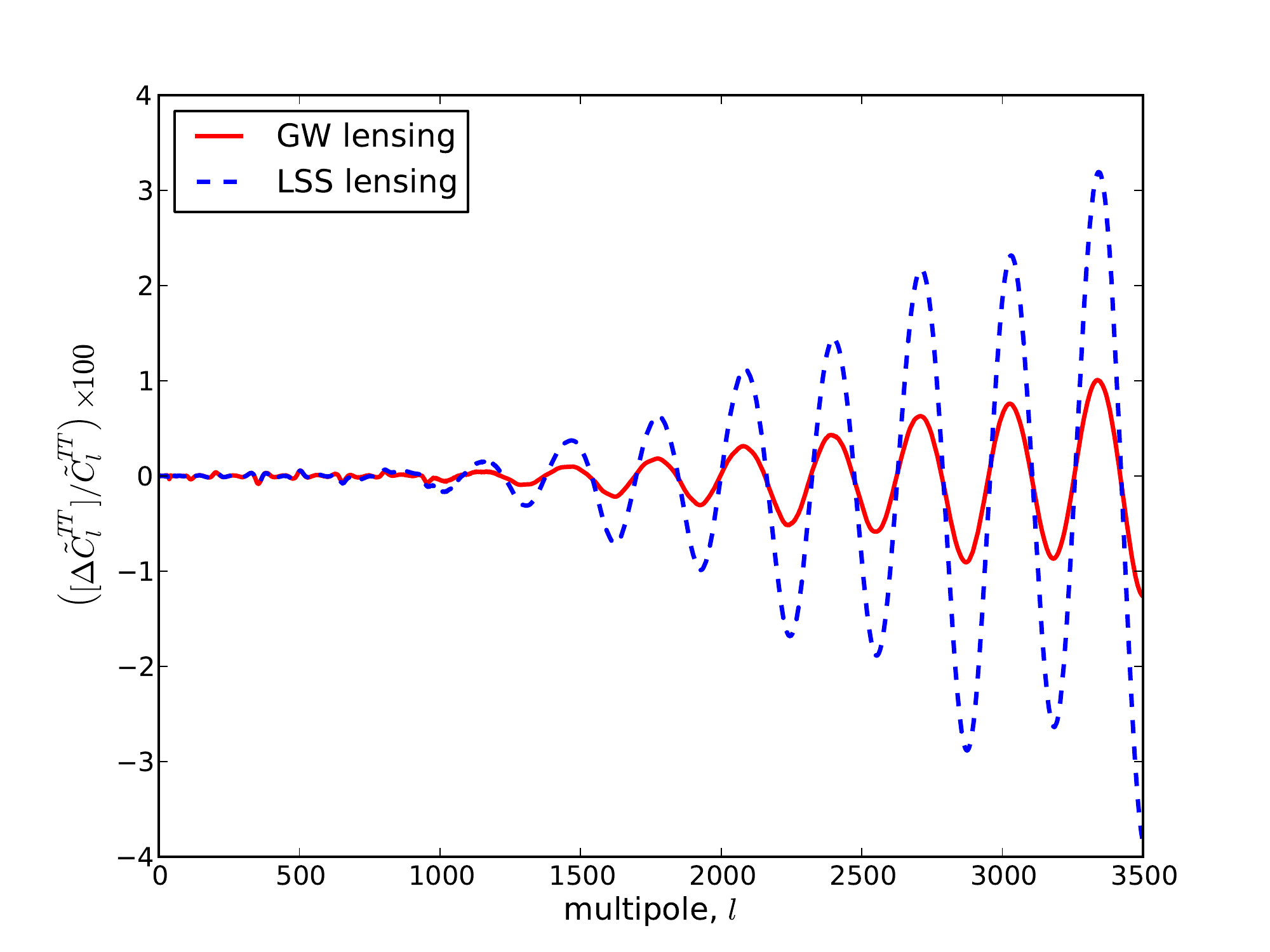}}
    \subfigure[$EE$]{\label{fig2b}  \includegraphics[height=6cm,width=8cm,angle=0]{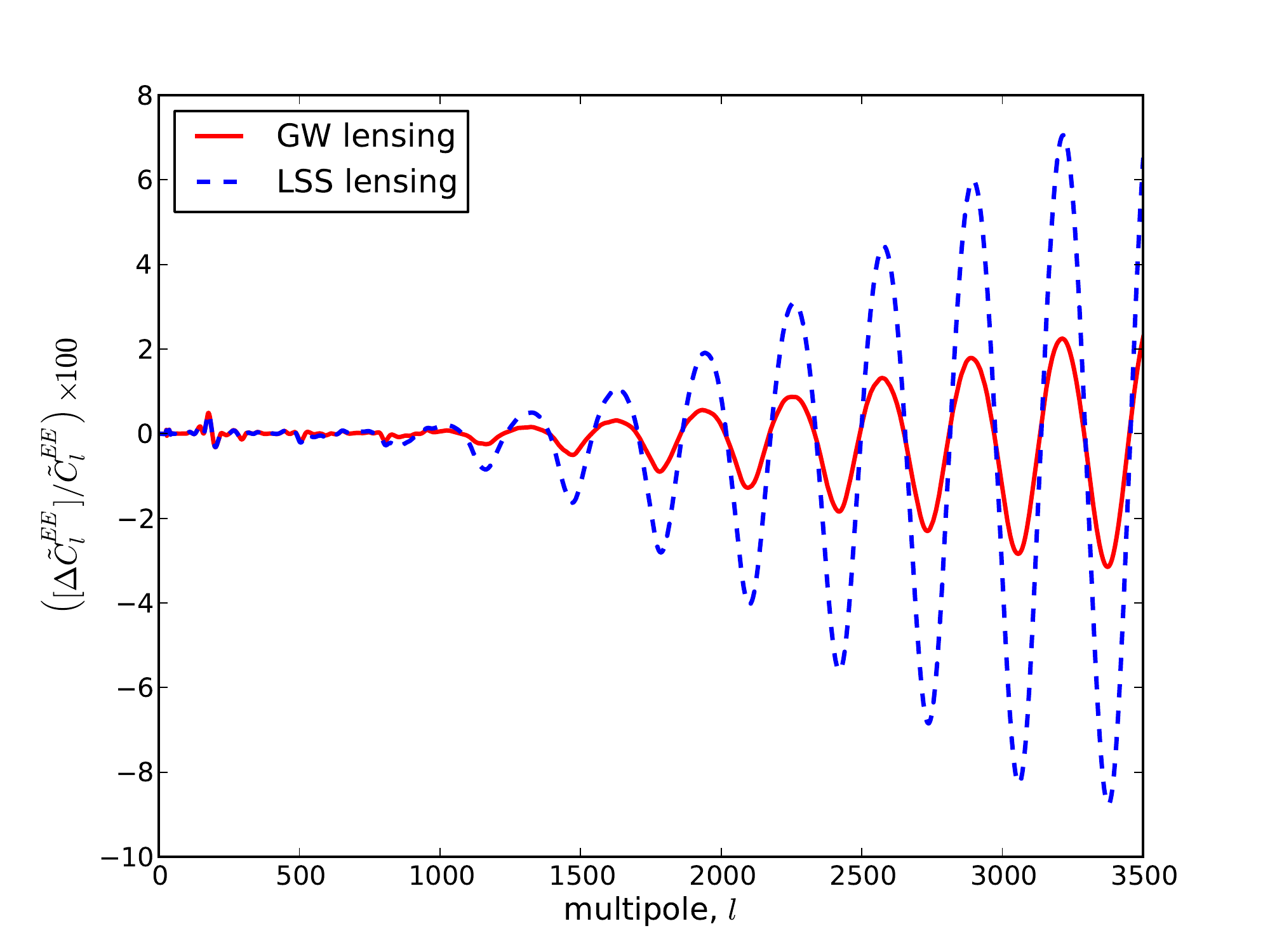}}
    \subfigure[$BB$]{\label{fig2c}  \includegraphics[height=6cm,width=8cm,angle=0]{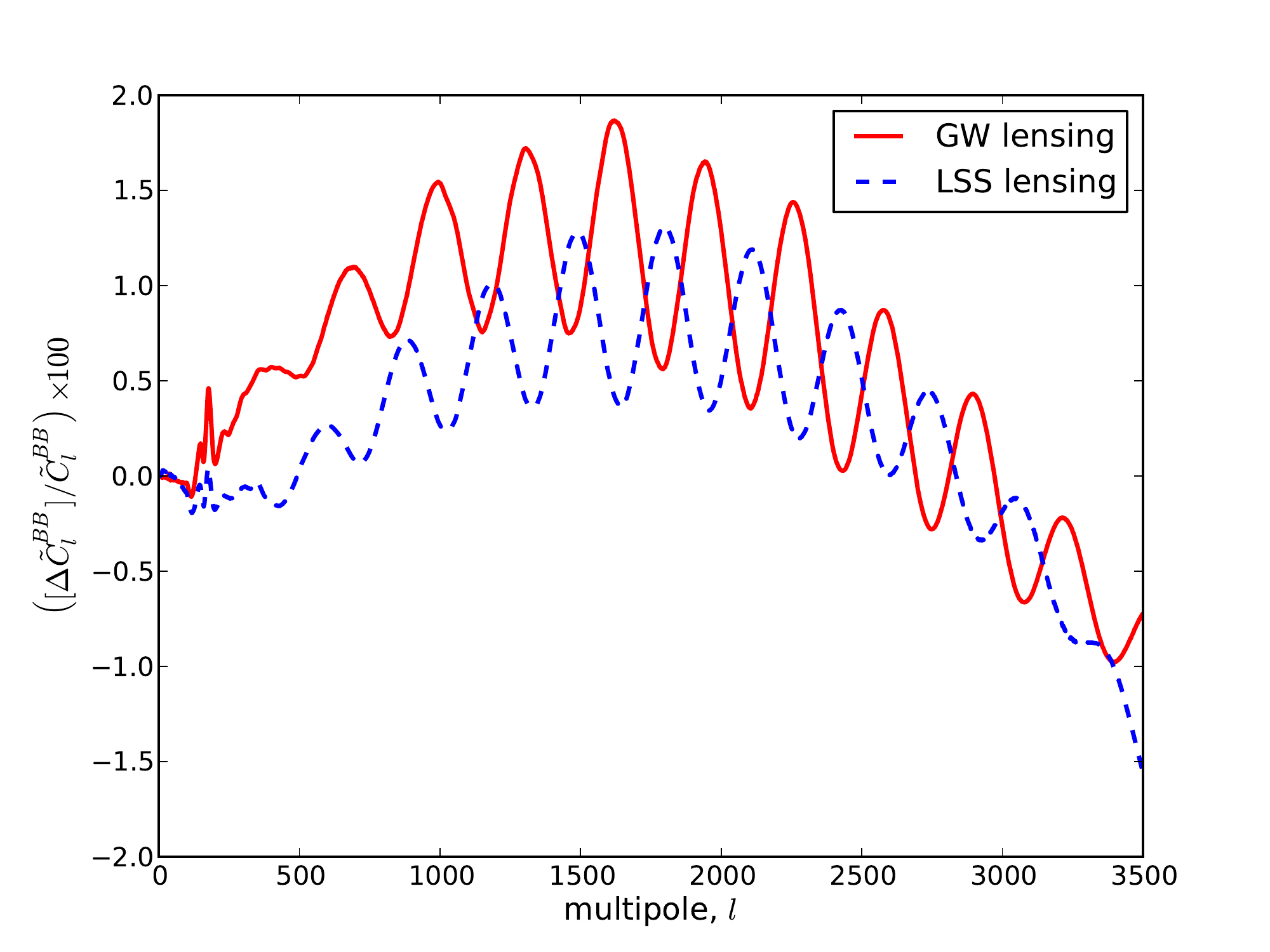}}
    \subfigure[$TE$]{\label{fig2d}  \includegraphics[height=6cm,width=8cm,angle=0]{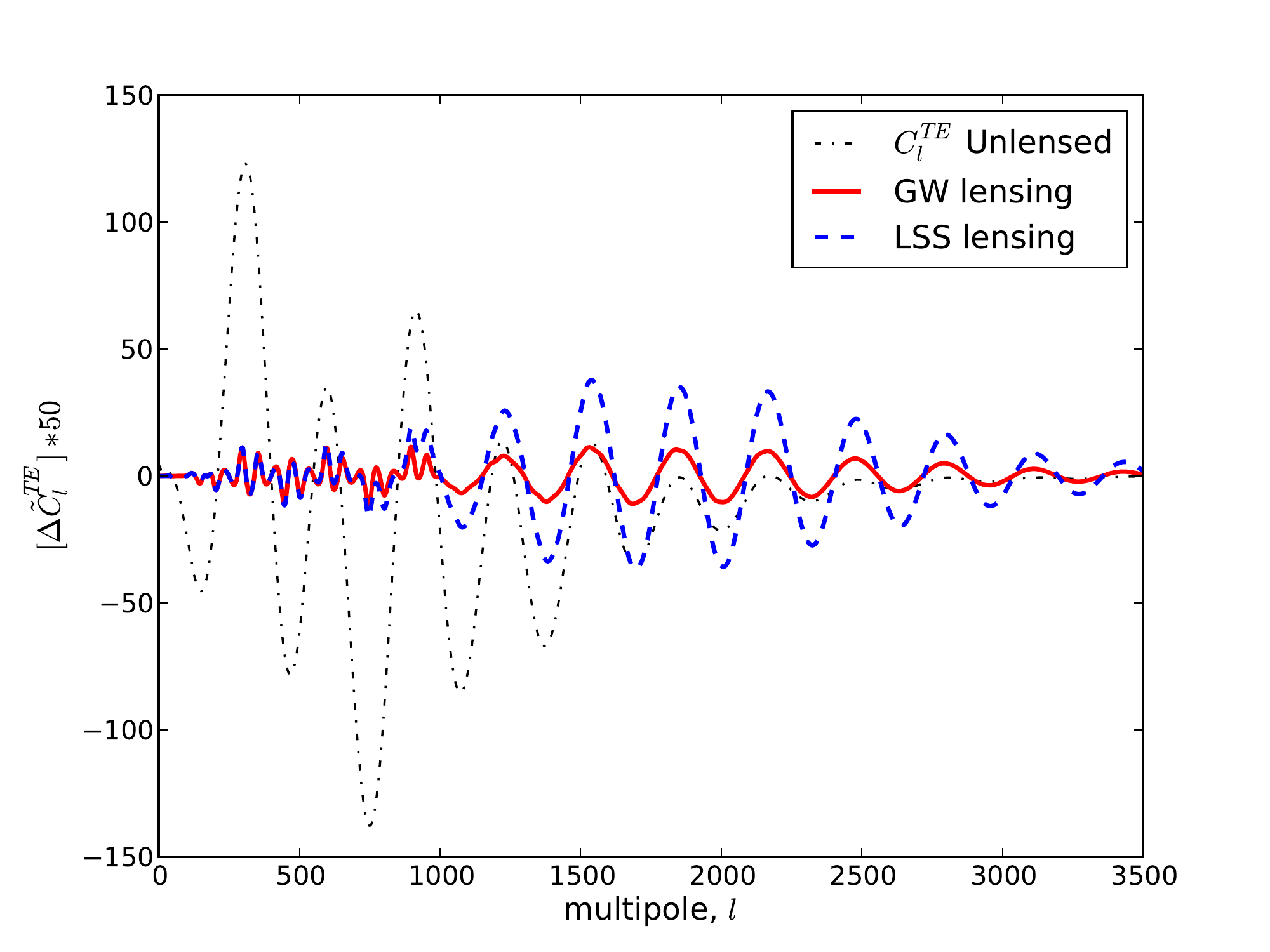}}
%	}
\end{center}
	\vskip -3mm
		\captionsetup{singlelinecheck=off,justification=raggedright}
	 \caption{The relative percentage difference between the lensed CMB spectra obtained by the harmonic space approach and the real space approach are plotted in Fig. \ref{fig2a} , \ref{fig2b} and \ref{fig2c}. For the $C_{l}^{TE}$ spectra, however, we plot the difference between the lensed spectra obtained by the two independent methods, magnified by a factor of 50. The unlensed $C_{l}^{TE}$ spectra is plotted along with it to roughly quantify the difference.}
	  \label{fig2}
\end{figure}
 
The relative accuracies of the lensed spectra obtained using these two independent methods is depicted in Fig. \ref{fig2}. It is observed that the difference between the lensed spectra obtained by the two independent methods is smaller for the case of lensing due to GW as compared to the case of lensing due to LSS, except in the case of the $C_{l}^{BB}$ spectrum where the trend is reversed.

This numerical exercise serves to establish two purposes. Firstly, it is seen that the lensed CMB spectra evaluated using these two different methods match each other very well. This establishes the correctness of the calculation of the lensing kernels and their equivalence. Secondly, it reconfirms the result that the real space kernels are more accurate than the harmonic space kernels at large multipoles. Fig.~\ref{fig2} indicates that the harmonic space method is accurate to within $1\%$ only upto a maximum multipole $\sim 1000$, beyond which the accuracy of this method diminishes very fast.
\newpage
\subsection{A comparative study of lensing due to LSS and GW.}
\label{results-b}
Since we set the deflection power spectra the same ( i.e $C_{l}^{\Omega\Omega}=C_{l}^{\psi\psi}$), the difference in the lensed spectra is only due to the difference between the lensing kernels $F^{\psi}$ and $F^{\Omega}$. Recall that these kernels have been shown to differ by a relative internal sign as discussed in Section~\ref{sec:similarity}.

\begin{figure}[!h]
\begin{center}
%	\mbox{
    \subfigure[$TT$]{\label{fig3a}  \includegraphics[height=6cm,width=8cm,angle=0]{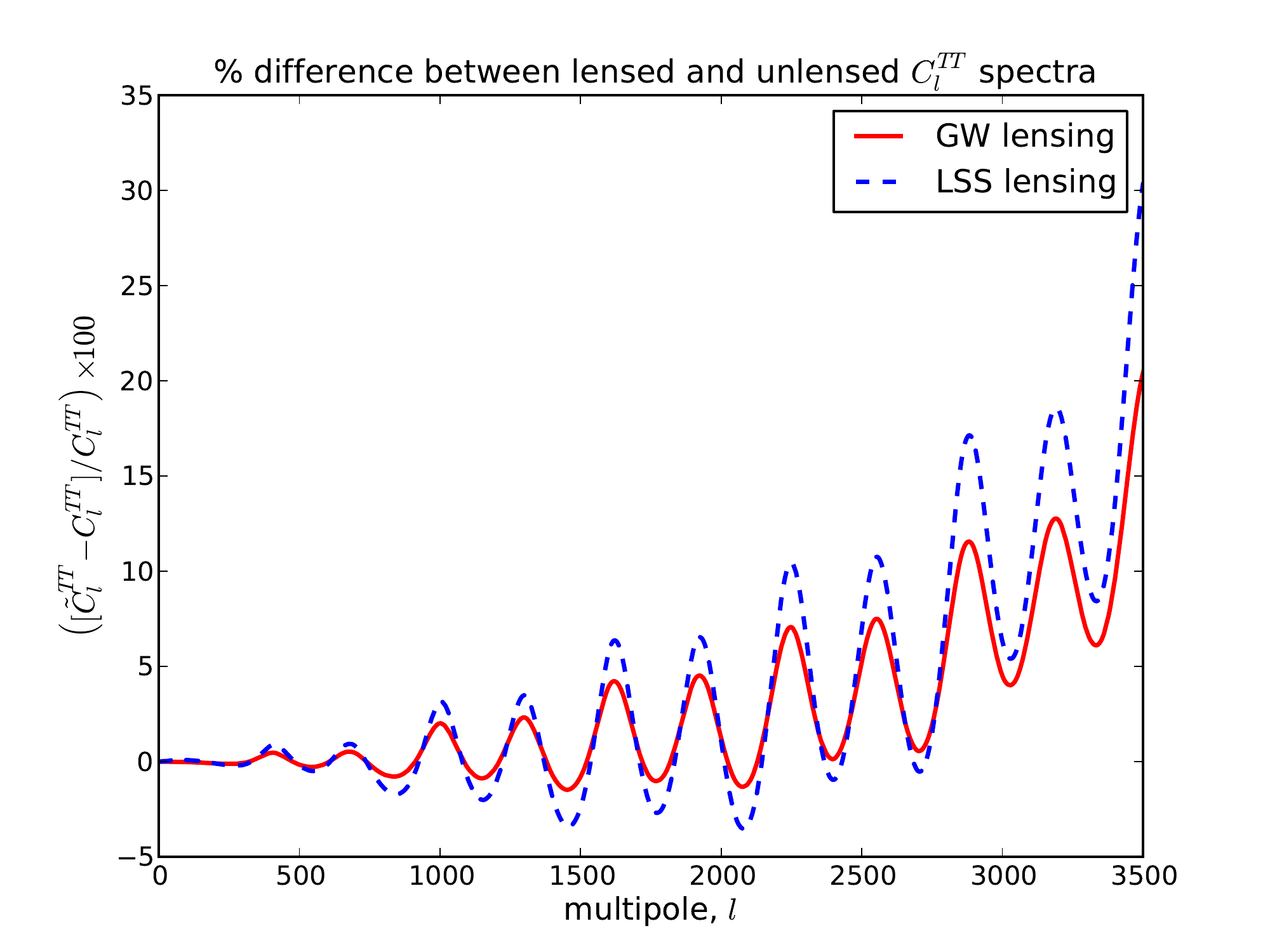}}
    \subfigure[$EE$]{\label{fig3b}  \includegraphics[height=6cm,width=8cm,angle=0]{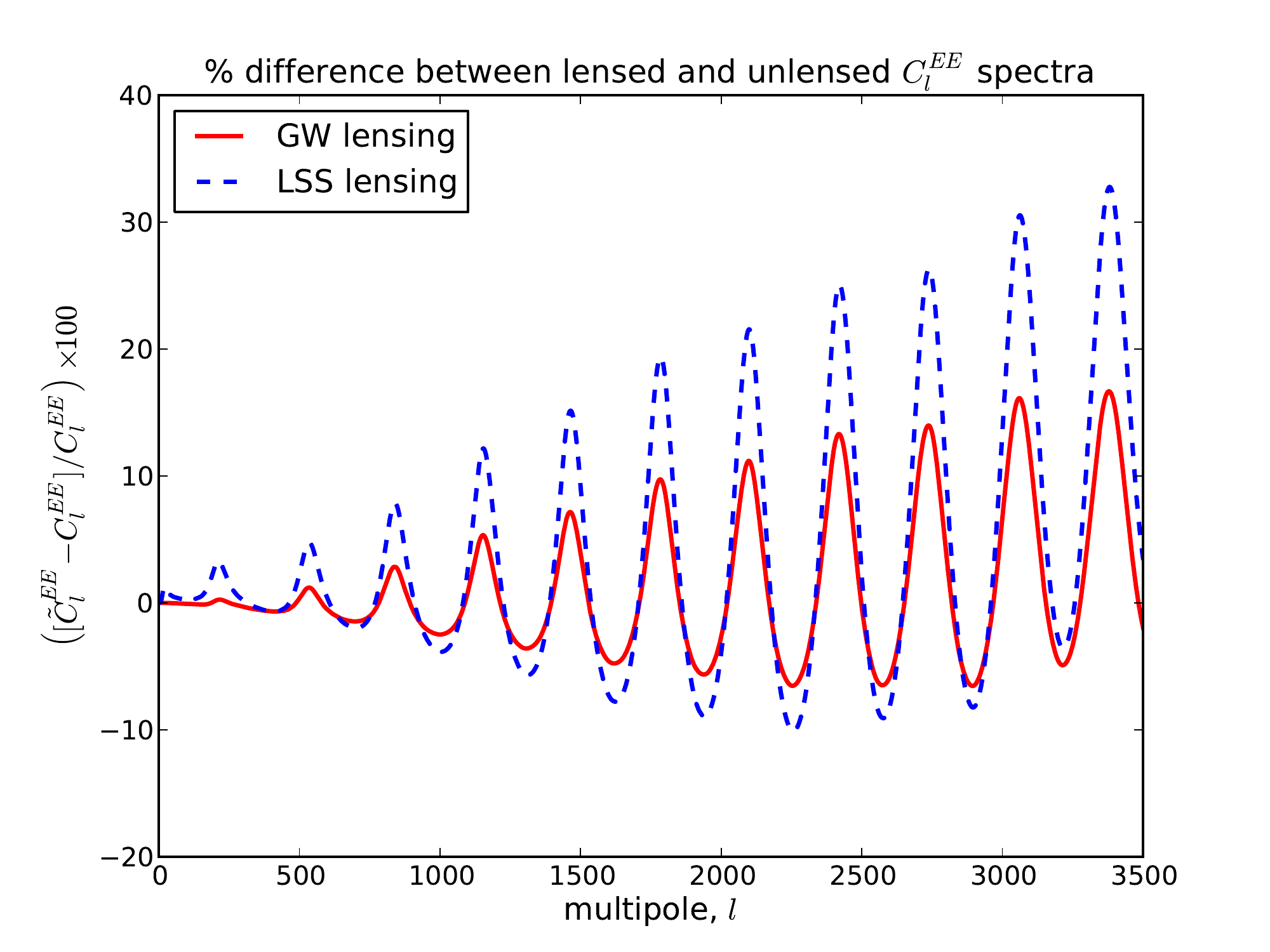}}
    \subfigure[$BB$]{\label{fig3c}  \includegraphics[height=6cm,width=8cm,angle=0]{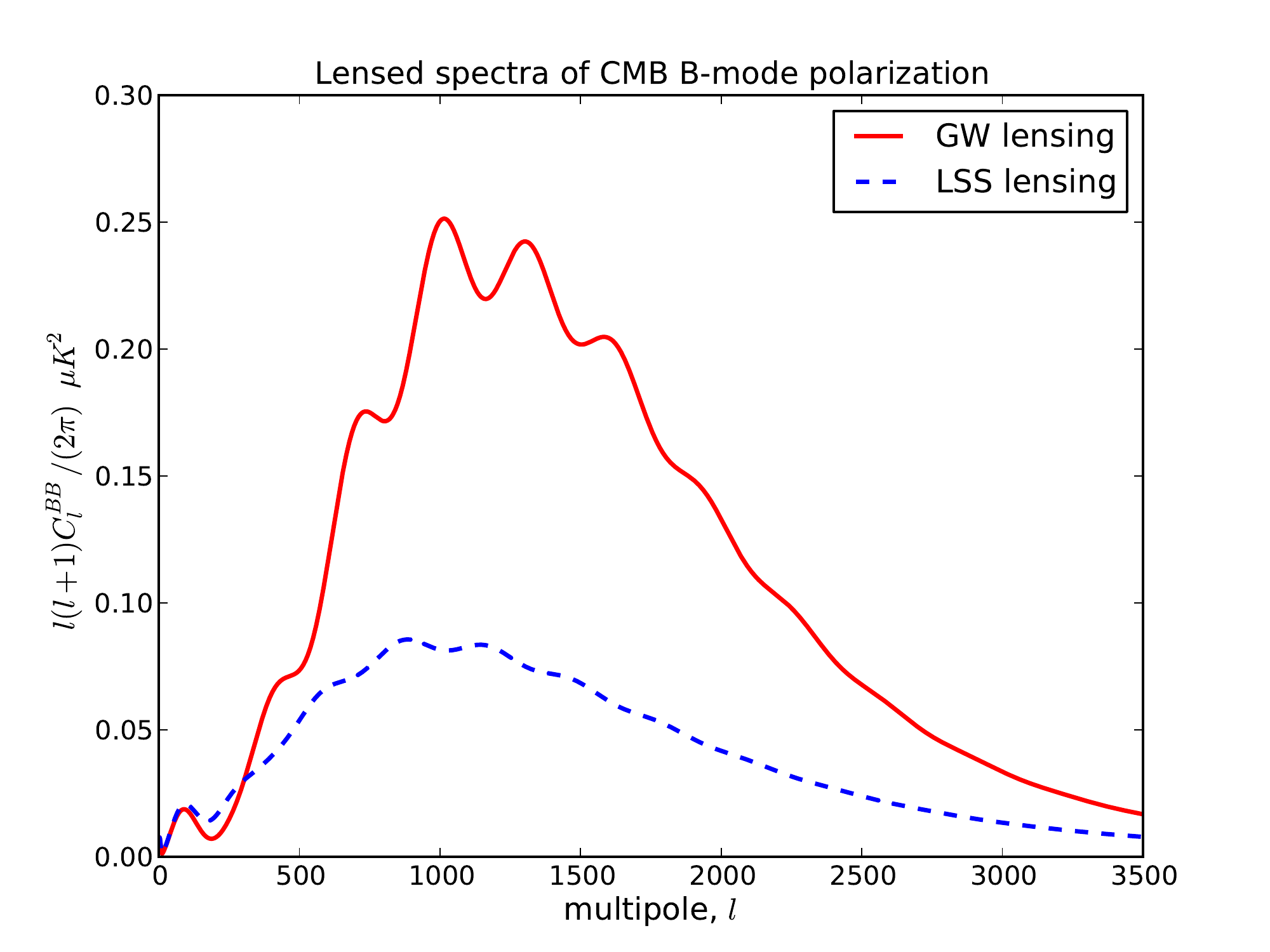}}
    \subfigure[$TE$]{\label{fig3d}  \includegraphics[height=6cm,width=8cm,angle=0]{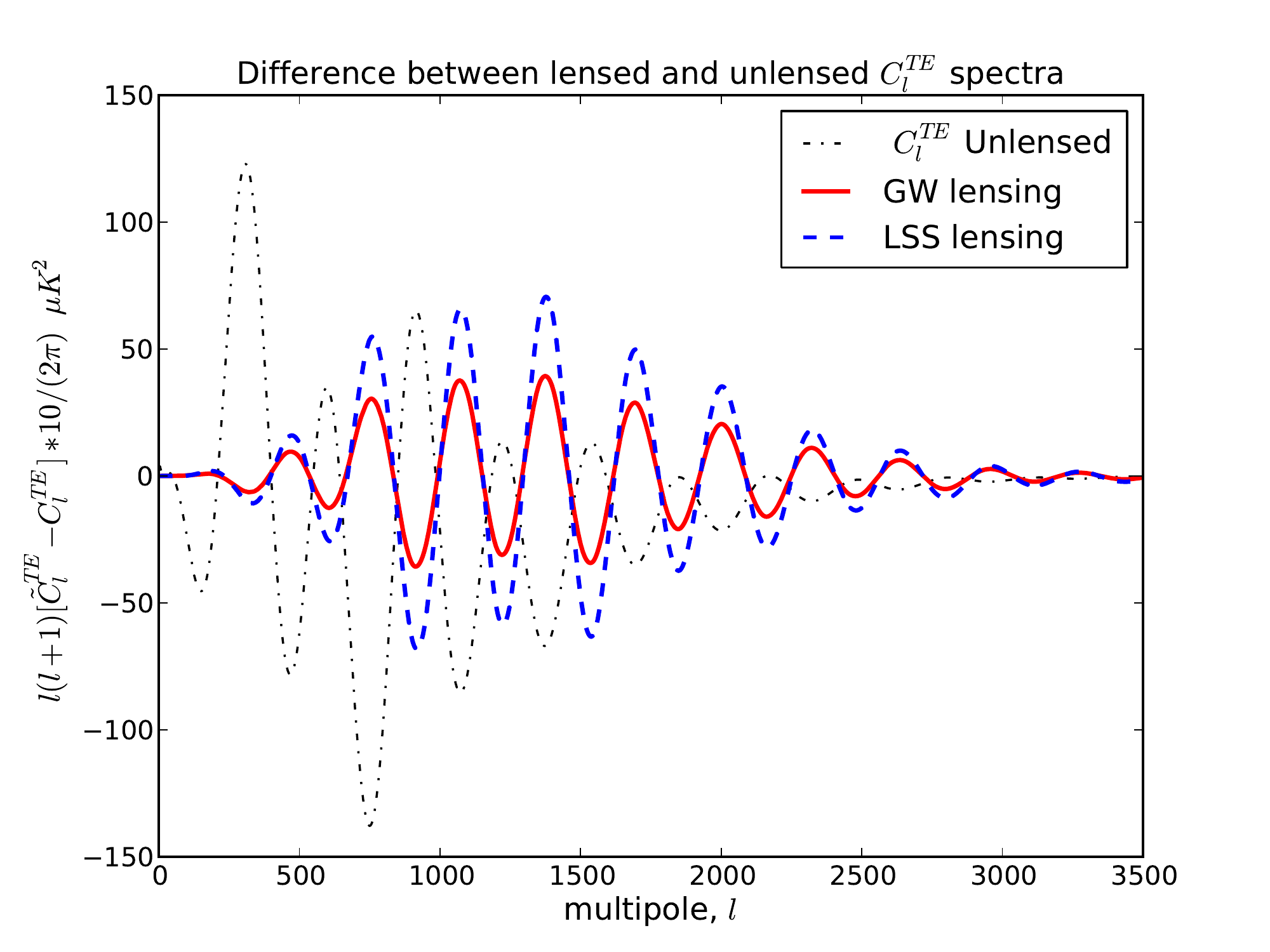}}
%	}
\end{center}
	\vskip -3mm
	\captionsetup{singlelinecheck=off,justification=raggedright}
	 \caption{Plotted in the Fig. \ref{fig3a} and \ref{fig3b} is the relative percentage difference between the lensed CMB spectra and the unlensed CMB spectra. For the case of $C_{l}^{BB}$ spectra, it is effectively null for large multipoles $(l > 200)$, the lensed spectrum itself quantifies the difference from the unlensed spectrum. For the $C_{l}^{TE}$ spectra (Fig. \ref{fig3d}), we plot the difference magnified by a factor of 10 between the lensed and unlensed spectra. The unlensed $C_{l}^{TE}$ spectra is plotted along with it to provide a reference for the level of the difference.}
	  \label{fig3}
\end{figure}

The resultant differences between the lensed CMB spectra are depicted in the figure below (Fig. \ref{fig3} ).
It is observed that the difference between the lensed and unlensed spectra is smaller for the case of lensing due to GW as compared to the case of lensing due to LSS, except in the case of the $C_{l}^{BB}$ spectrum where the trend is reversed. This indicates that even though lensing by GW is not as effective in distorting the $C_{l}^{TT}$ spectra as compared to lensing by LSS, it is more efficient in mediating the power transfer between the $E$-mode and $B$-mode of CMB polarization. 

The $C_{l}^{EE}$ spectrum is less distorted in the case of lensing by GW as compared to lensing by LSS,
 but there is more power transfer to the $B$-mode of CMB polarization in the former case, as seen in Fig.~\ref{fig3b} and Fig.~\ref{fig3c}. We have tested that the total amount of polarization is conserved. 
Further, these lensed spectra evaluated using two different methods and have been shown to be consistent with each other, which validates the consistency and correctness of our calculations and numerical verifications.

\section{Conclusions and Outlook}

While the lensing due to scalar, or large-scale structure perturbations has been studied fairly well in the past, the lensing by tensor perturbations has not been as extensively explored. In this paper we have been able to provide a full, detailed treatment of the latter, both analytically and numerically. In so doing, we reveal striking similarities in the structure of the lensing kernels for the case of lensing by scalar and tensor perturbations, which may have possible connections to parity invariance of scalar, in contrast to tensor perturbations. These similarities carry through for the cases of the lensing kernels for the polarization $E$ and $B$ mode fluctuations as well. Also, we have been able to extend the analysis  to the real space correlation function method for the case of lensing due to tensor perturbations (gravitational waves). We have shown that the two independent methods, viz. the real-space correlation function approach, as well as the spherical harmonic-space full sky approach lead to \textit{identical numerical results}, when an important modification is made in the expressions for the lensed polarization and cross-correlation power spectra derived previously in the literature. 

\begin{figure}
\includegraphics[height=8cm,width=10cm,angle=0]{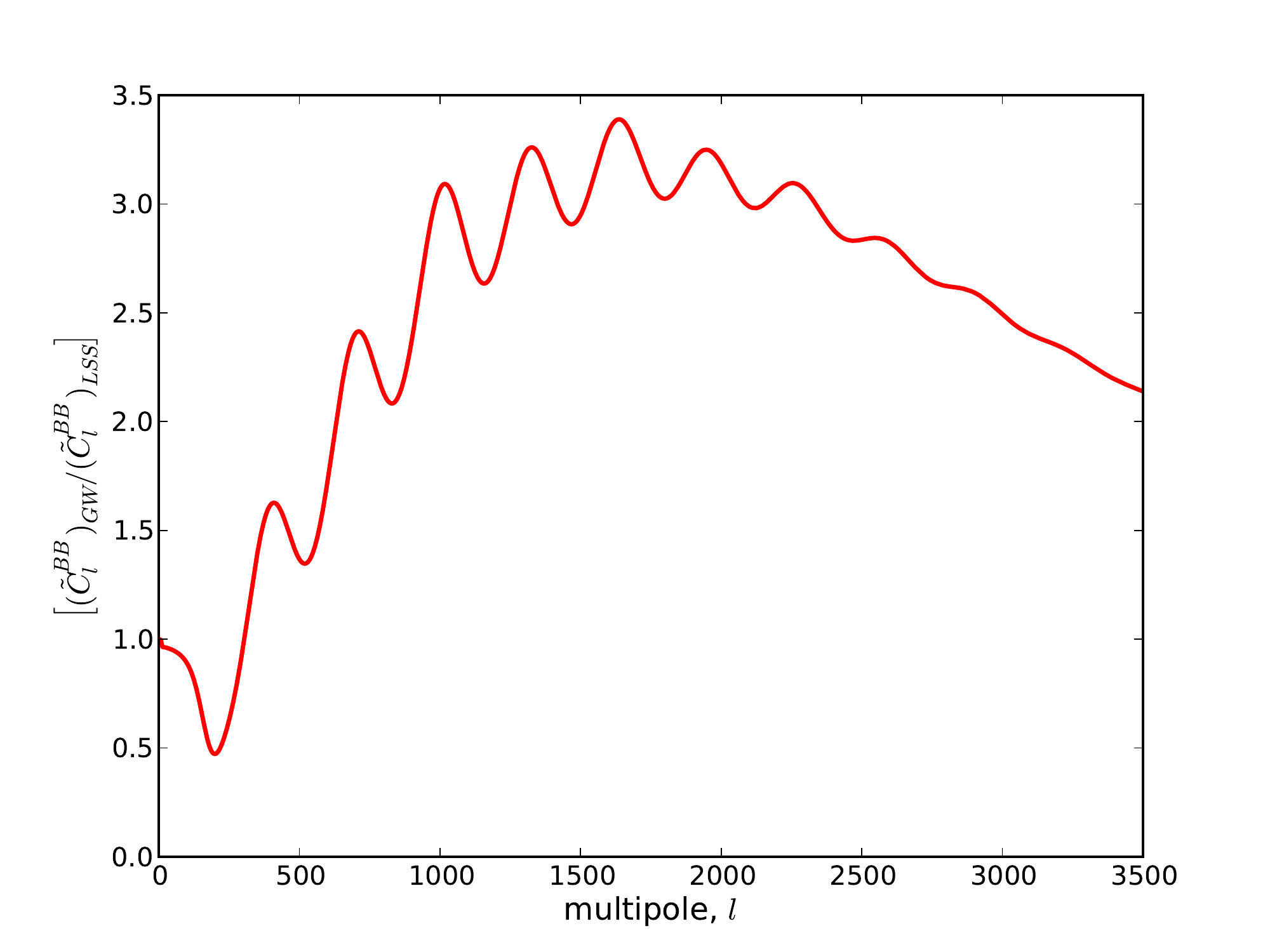} 
\caption{This plot shows the ratio of the lensed $B$-mode spectrum of CMB polarization due to GW to lensed $B$-mode spectrum of CMB polarization due to LSS.  It is seen that at large multipoles $(l > 300)$, where effectively all the $B$-mode power is due to lensing, GW are more efficient at transferring power to the $B$-mode of CMB polarization.}
\label{BBefficiency}
\end{figure}

It is well known that lensing results in a mixing of power between the $E$-mode and $B$-mode of CMB polarization. Through the numerical studies on the lensing kernels we have quantitatively established that, this mixing of power is more efficiently mediated by the curl type photon displacements which are primarily induced due to GW. In Figure.~\ref{BBefficiency} we plot the ratio of lensed $B$-mode spectra due to LSS and GW. This indicates that the $B$-mode of CMB polarization can be used to place interesting constraints on GW energy densities \cite{arandts2011}. 

There are several directions in which this work can be taken further. To begin with, it would be interesting to provide an analytical explanation for the similarities and differences between the lensing kernels for scalar and tensor perturbations. Also, a likely connection between parity properties and the relative sign flip between the two terms in the harmonic space analysis, is worth investigating. In the near future, missions such as CMBPol\cite{cmbpol} and ACTPol\cite{actpol} promise to place interesting constraints on the primordial gravitational wave power spectrum as also possibly detect the primordial $B$-mode power spectra. This has implications for inflationary models and high energy particle physics. The weak lensing of the CMB by stochastic gravitational waves mixes power between the $E$ and $B$ modes and hence, modelling this lensing effect helps in the separation of the primordial $B$-mode power spectrum from the effect of leakage due to lensing from the $E$-mode to the $B$-mode. 

\section*{Acknowledgments}
We acknowledge useful discussions with Sanjit Mitra. HP acknowledges support by the Shyama Prasad Mukherjee Fellowship of the Council of Scientific and Industrial Research (CSIR), India. AR acknowledges support from CSIR (Grant award no. 20-6/2008(II)E.U.-IV). TS acknowledges support from the Swarnajayanti fellowship, DST, India.

%\input{./sections/appendices}
%%%%%%%%%%%%%%%%%%%%%%%%%%%%%%%%%%%%%%%%%%%%%%%%%%%%%%%%%%
\section*{Appendices}
\appendix
%%%%%%%%%%%%%%%%%%%%%%%%%%%%%%%%%%%%%%%%%%%%%%%%%%%%%%%%%%
\section{Similarity of scalar and tensor lensing kernels}
\label{app:polharspace2}

In all of lensing literature, the LSS lensing kernels for CMB polarization spectra are expressed in a mathematical form very different from the corresponding kernels evaluated for the case of GW lensing \cite{cooray1}. Here we derive the LSS polarization lensing kernel and express it in a form which looks similar to the GW polarization lensing kernel. 

Evaluating the LSS lensing kernel for CMB polarization requires the evaluation of the following integral,
\begin{equation}
_{2}I_{lml_1m_1l_2m_2}^\psi =\int d\mathbf{\hat{n}} \  _{2}Y_{lm}^* \nabla_a Y_{l_1m_1} \nabla^a _{2} Y_{l_2m_2} \,.                                                                                                                         
\label{lss-lens-pol-int}
\end{equation} 
Usually the above integral is solved by integrating by parts which leads to the form of the lensing kernels expressed in its usual form \cite{waynehu}.

We use a different method to evaluate the above integral, which allows us to express the LSS lensing kernel and GW lensing kernels in similar mathematical forms. We make use of the following mathematical identity,
\begin{equation}
 \nabla_a ({}_s Y_{lm}) = \sqrt{\frac{(l - s)(l + s +1)}{2}} {}_{s+1} Y_{lm} (\mathbf{m}_+)_a - \sqrt{\frac{(l + s)(l - s +1)}{2}} {}_{s-1} Y_{lm} (\mathbf{m}_-)_a  \, ,
 \end{equation}
 where $\mathbf{m}_+$ and $\mathbf{m}_-$ are related to the standard basis functions on the sphere, $\mathbf{e}_{\theta}$ and $\mathbf{e}_{\phi}$, through the following expression,
\begin{equation}
 \mathbf{m}_{\pm} = \frac{1}{\sqrt{2}} (\mathbf{e}_{\theta} \mp  i\mathbf{e}_{\phi}) \, .
\end{equation} 

\noindent Expanding the gradients in \eq{lss-lens-pol-int} using the above identity allows us to express the integral in the following form,
\begin{eqnarray}
_{2}I_{lml_1m_1l_2m_2}^\psi & =& \int d\mathbf{\hat{n}} \ _{2}Y_{lm}^* \\ &\times & \left[\sqrt{\frac{l_1(l_1+1)}{2}} {}_1 Y_{l_1m_1} (\mathbf{m}_+)_a -  \sqrt{\frac{l_1(l_1+1)}{2}} {}_{-1} Y_{l_1m_1} (\mathbf{m}_-)_a  \right] \nn
 &\times & \Bigg[\sqrt{\frac{(l_2 - 2)(l_2+3)}{2}} {}_3 Y_{l_2m_2} (\mathbf{m}_+)_a -\sqrt{\frac{(l_2 +2)(l_2-1)}{2}} {}_{1} Y_{l_2m_2} (\mathbf{m}_-)_a  \Bigg]  \,.  \nonumber 
\label{ipsigradexpand}
\end{eqnarray} 

\noindent Further, on using the identities $(\mathbf{m}_+)_a (\mathbf{m}_+)_a = 0$ and $(\mathbf{m}_+)_a (\mathbf{m}_-)_a = 1$, the integral can be reduced to the following simple form,
\begin{eqnarray}
_{2}I_{lml_1m_1l_2m_2}^\psi &=& -\frac{1}{2}\int d\mathbf{\hat{n}} \ (_{2}Y_{lm}^*) ({}_1 Y_{l_1m_1})({}_{1} Y_{l_2m_2}) \sqrt{l_1(l_1 + 1)(l_2 +2)(l_2 -1)} \nonumber\\
&-&\frac{1}{2}\int d\mathbf{\hat{n}} \ (_{2}Y_{lm}^*) ({}_{-1} Y_{l_1m_1})({}_{3} Y_{l_2m_2}) \sqrt{l_1(l_1 + 1)(l_2 -2)(l_2 +3)} \,. \nn
\label{simplified-lss-lens-int}
\end{eqnarray}

\noindent The standard integrals involved in the above equation are known to have the following solution \cite{varshalovich},
\begin{eqnarray}
 \int d\mathbf{\hat{n}}
({}_{s_1}Y_{l_1m_1}^*)({}_{s_2}Y_{l_2m_2})({}_{s_3}Y_{l_3m_3}) &=&
\quad (-1)^{m_1+s_1}
\frac{\Pi_{l_1} \Pi_{l_2} \Pi_{l_3}}{\sqrt{4\pi}} \\
&\times &
\begin{pmatrix}l_1&l_2&l_3\\
s_1&-s_2&-s_3\end{pmatrix}
\begin{pmatrix}l_1&l_2&l_3\\-m_1&m_2&m_3\end{pmatrix} \nonumber \,,
\label{threesphar}
\end{eqnarray}
where $\Pi_l = \sqrt{2l+1}$.
The above solution is non-vanishing only when the following condition is satisfied $s_1 = s_2 + s_3$. Using these facts, the integral in \eq{simplified-lss-lens-int} can be simplified to the following form,
\begin{eqnarray}
&_{2}I_{lml_1m_1l_2m_2}^\psi &= -(-1)^m \sqrt{\frac{l_1 (l_1 + 1)}{2}}\frac{\Pi_{l}\Pi_{l_1} \Pi_{l_2}}{\sqrt{4\pi}}  \begin{pmatrix}l&l_1&l_2\\
-m&m_1&m_2\end{pmatrix} \\
&& \left[ \mathcal{C}_1 \begin{pmatrix}l&l_1&l_2\\
2&-1&-1\end{pmatrix}
 + \mathcal{C}_2 \begin{pmatrix}l&l_1&l_2\\
2&1&-3\end{pmatrix}\right]\nonumber \,,
\end{eqnarray} 
where $\mathcal{C}_1=\sqrt{\frac{(l_2 + 2)(l_2 -1)}{2}}$ and $\mathcal{C}_2=\sqrt{\frac{(l_2 -2)(l_2 +3)}{2}} $ . 

The LSS lensing kernel associated with lensing of CMB polarization is related to $_{2}I_{lml_1m_1l_2m_2}^\psi $ through the following equation,
\begin{equation}
|_{2}F_{ll_1l_2}^\psi|^2 = (2l +1) \sum_{m_1 m_2} |_{2}I_{lml_1m_1l_2m_2}^\psi|^2 .
\end{equation} 

Using the mathematical identity for Clebsch-Gordon coefficients, 
\begin{equation}
 \sum_{m_1 m_2} \begin{pmatrix}l&l_1&l_2\\
-m&m_1&m_2\end{pmatrix} \begin{pmatrix}l&l_1&l_2\\
-m&m_1&m_2\end{pmatrix}  = \frac{1}{2 l +1} \,,
\end{equation} 
it can be shown that the lensing kernel is given by the following expression,
\begin{eqnarray}\label{lss-kernel}
 _{2}F_{ll_1l_2}^\psi  &=& \sqrt{\frac{l_1 (l_1 + 1)}{2}}\frac{\Pi_{l}\Pi_{l_1} \Pi_{l_2}}{\sqrt{4\pi}} 
\left[ \mathcal{C}_1 \begin{pmatrix}l&l_1&l_2\\
2&-1&-1\end{pmatrix}
 + \mathcal{C}_2 \begin{pmatrix}l&l_1&l_2\\
2&1&-3\end{pmatrix}\right]\nonumber\,.\\
\end{eqnarray} 
This proves the result in \eq{2fpsiequiv}.

Next we derive the corresponding GW lensing kernel. In this case we need to evaluate the following integral,
\begin{equation}
_{2}I_{lml_1m_1l_2m_2}^\Omega =\int d\mathbf{\hat{n}} \ \epsilon^b_a \  _{2}Y_{lm}^* \nabla^a {}_{2} Y_{l_2m_2} \nabla_b  Y_{l_1m_1} \,,                                                                                                                 
\label{gw-lens-pol-int}
\end{equation} 
where $\epsilon^b_a$ is the two dimensional completely antisymmetric Levi-Civita symbol.

Expanding the gradients in the above equation, it can be rewritten in the following form,
\begin{eqnarray}
_{2}I_{lml_1m_1l_2m_2}^\Omega &=& \int d\mathbf{\hat{n}} \ \epsilon^b_a \ _{2}Y_{lm}^* \\ &\times &\left[\sqrt{\frac{l_1(l_1+1)}{2}} {}_1 Y_{l_1m_1} (\mathbf{m}_+)_b -  \sqrt{\frac{l_1(l_1+1)}{2}} {}_{-1} Y_{l_1m_1} (\mathbf{m}_-)_b  \right] \nn
 & \times & \left[\sqrt{\frac{(l_2 - 2)(l_2+3)}{2}} {}_3 Y_{l_2m_2} (\mathbf{m}_+)^a -  \sqrt{\frac{(l_2 +2)(l_2-1)}{2}} {}_{1} Y_{l_2m_2} (\mathbf{m}_-)^a  \right]  \,.  \nonumber                                                                                                           
 \end{eqnarray} 
 
Using the identities $(\mathbf{m}_+)_a (\mathbf{m}_-)_b \epsilon^{ab} = i$ and $(\mathbf{m}_-)_a (\mathbf{m}_+)_b \epsilon^{ab} = -i$, along with the standard result for integration over three spherical harmonics (\eq{threesphar}), the integral in \eq{gw-lens-pol-int} can be solved to give the following,
\begin{eqnarray}
_{2}I_{lml_1m_1l_2m_2}^\Omega &=& -i (-1)^m \sqrt{\frac{l_1 (l_1 + 1)}{2}}\frac{\Pi_{l}\Pi_{l_1} \Pi_{l_2}}{\sqrt{4\pi}}  \begin{pmatrix}l&l_1&l_2\\
-m&m_1&m_2\end{pmatrix} \\
&& \left[ \mathcal{C}_1 \begin{pmatrix}l&l_1&l_2\\
2&-1&-1\end{pmatrix}
 - \mathcal{C}_2 \begin{pmatrix}l&l_1&l_2\\
2&1&-3\end{pmatrix}\right] \nonumber \,,
\end{eqnarray} 
where $\mathcal{C}_1=\sqrt{\frac{(l_2 + 2)(l_2 -1)}{2}}$ and $\mathcal{C}_2=\sqrt{\frac{(l_2 -2)(l_2 +3)}{2}} $.

Following the same procedure as the one used to arrive at the LSS lensing kernel, it can be shown that the GW lensing kernel is given by the following expression,
\begin{eqnarray} \label{gw-kernel}
 _{2}F_{ll_1l_2}^\Omega  &=& \sqrt{\frac{l_1 (l_1 + 1)}{2}}\frac{\Pi_{l}\Pi_{l_1} \Pi_{l_2}}{\sqrt{4\pi}} 
\left[ \mathcal{C}_1 \begin{pmatrix}l&l_1&l_2\\
2&-1&-1\end{pmatrix}
 - \mathcal{C}_2 \begin{pmatrix}l&l_1&l_2\\
2&1&-3\end{pmatrix}\right]\nonumber\,.\\
\end{eqnarray} 

Comparing \eq{lss-kernel} with \eq{gw-kernel} reveals that the two kernels only differ in the relative sign which appears between the two terms.

Next we explicitly evaluate the following integral,
\begin{equation}
_{-2}I_{lml_1m_1l_2m_2}^\Omega =\int d\mathbf{\hat{n}} \ \epsilon^b_a \ _{-2}Y_{lm}^* \nabla^a {}_{-2} Y_{l_2m_2} \nabla_b  Y_{l_1m_1}                                                                                                                        \,.
\end{equation}  
Following the same procedure as previously described, it can be shown that this integral evaluates to the following expression,
\begin{eqnarray}
_{2}I_{lml_1m_1l_2m_2}^\Omega &=& i (-1)^m \sqrt{\frac{l_1 (l_1 + 1)}{2}}\frac{\Pi_{l}\Pi_{l_1} \Pi_{l_2}}{\sqrt{4\pi}}  \begin{pmatrix}l&l_1&l_2\\
-m&m_1&m_2\end{pmatrix} \\
&& \left[ \mathcal{C}_1 \begin{pmatrix}l&l_1&l_2\\
-2&1&1\end{pmatrix}
 - \mathcal{C}_2 \begin{pmatrix}l&l_1&l_2\\
-2&-1&3\end{pmatrix}\right]\nonumber \,,
\end{eqnarray} 
where $\mathcal{C}_1=\sqrt{\frac{(l_2 + 2)(l_2 -1)}{2}}$ and $\mathcal{C}_2=\sqrt{\frac{(l_2 -2)(l_2 +3)}{2}} $.

Using the following mathematical identity, 
\begin{eqnarray}
\begin{pmatrix}l_1&l_2&l_3\\m_1&m_2&m_3\end{pmatrix} = (-1)^{l_1 + l_2 + l_3} \begin{pmatrix}l_1&l_2&l_3\\-m_1&-m_2&-m_3\end{pmatrix} \,,
\label{wignerreln}
\end{eqnarray} 
it can be shown that,
\begin{equation}
_{-2}I_{lml_1m_1l_2m_2}^\Omega =  (-1)^{l + l_1 +l_2+1} {}_{2}I_{lml_1m_1l_2m_2}^\Omega \, .
\label{ilmplminus}
\end{equation} 

For similar integrals (\eq{lss-lens-pol-int}) encountered during evaluation of the LSS lensing kernel, it can be similarly shown that,
\begin{equation}
_{-2}I_{lml_1m_1l_2m_2}^\psi =  (-1)^{l + l_1 +l_2} {}_{2}I_{lml_1m_1l_2m_2}^\psi \, .
\label{ilmplminus-lss}
\end{equation}
%%%%%%%%%%%%%%%%%%%%%%%%%%%%%%%%%%%%%%%%%%%%%%%%%%%%%%%%%%
\section{Lensed polarization power spectra}
\label{app:polharspace1}
As described in Refs. \cite{waynehu, cooray1}, we begin by expanding the polarization field in a Taylor series and perform the spherical harmonic transform of the resulting expression to arrive at the following harmonic space equation,
\begin{eqnarray}
&& {}_\pm\tilde{a}_{lm}^X = {}_\pm
a_{lm}^X - \sum_{l_1m_1l_2m_2} \Big[{}_\pm a_{l_2m_2}^X \Big(
{}_{\pm 2}I^\psi_{lml_1m_1l_2m_2}\psi_{l_1m_1}
 + {}_{\pm 2} I^\Omega_{lml_1m_1l_2m_2}\Omega_{l_1m_1}\Big)\Big]
 \nonumber\\ 
 &+&\frac{1}{2}\sum_{l_1m_1l_2m_2l_3m_3} \Big[{}_\pm
a_{l_2m_2}^X \Big({}_{\pm
2}J^\psi_{lml_1m_1l_2m_2l_3m_3}\psi_{l_1m_1}\psi^{
*}_{l_3m_3}  \nonumber\\
&+& {}_{\pm
2}J^\Omega_{lml_1m_1l_2m_2l_3m_3}\Omega_{l_1m_1}\Omega^{
*}_{l_3m_3}\Big)\Big] \, ,
\end{eqnarray}
with the explicit form of the integrals given by the following expressions,
\begin{eqnarray}
 _{\pm 2}I_{lml_1m_1l_2m_2}^\psi &=&\int d\mathbf{\hat{n}} \ _{\pm 2}Y_{lm}^* \nabla_a Y_{l_1m_1} \nabla^a _{\pm 2} Y_{l_2m_2}\nonumber\\
_{\pm 2} I_{lml_1m_1l_2m_2}^\Omega &=&\int  d\mathbf{\hat{n}}  \ _{\pm 2} Y_{lm}^* \nabla^a _{\pm 2}Y_{l_2m_2} \nabla_b Y_{l_1m_1}\epsilon^b{}_a \nonumber\\
_{\pm 2}J_{lml_1m_1l_2m_2l_3m_3}^\psi &=& \int d\mathbf{\hat{n}} \
_{\pm 2}Y_{lm}^* \nabla^a \nabla^b _{\pm 2}Y_{l_2m_2} \nabla_a Y_{l_1m_1} \nabla_b  Y_{l_3m_3}^* \nonumber \\
_{\pm 2} J_{lml_1m_1l_2m_2l_3m_3}^\Omega &=& \int d\mathbf{\hat{n}} \
_{\pm 2}Y_{lm}^*\nabla^a \nabla^b _{\pm 2}Y_{l_2m_2} \nabla_c Y_{l_1m_1} \nabla_d Y_{l_3m_3}^*\epsilon^c{}_a\epsilon^d{}_b
\, . 
\end{eqnarray}

First we consider the case of lensing due to LSS ($\Omega = 0$), in which case we get the following expression for the lensed spectrum of the $E$-mode of polarization,
\begin{eqnarray}
 \tilde{C}_l^{EE} &=& \frac{(\langle {}_{+} a_{lm}^{\tilde{X}} {}_{+} a_{lm}^{*\tilde{X}} \rangle + \langle {}_{+} a_{lm}^{\tilde{X}} {}_{-} a_{lm}^{*\tilde{X}} \rangle)}{2} \nonumber\\
 &=& C_{l}^{EE}+\frac{1}{2}\sum_{l_1l_2}C_{l_1}^
{\psi} {}_{2} S_1^{\psi}\left[(C_{l_2}^{EE}+C_{l_2}^{BB})+(-1)^L
(C_{l_2}^{EE}-C_{l_2}^{BB})\right]\nonumber\\
&+& C_l^{EE} \sum_{l_1} C_{l_1}^{\psi}{}_{2}S_2^{\psi}     
\label{cllensedeelss}
\end{eqnarray} 
where $L \equiv l + l_1 + l_2, \ {}_{2}S_1^\psi = \sum_{m_1m_2}|{}_{\pm 2}I_{lml_1m_1l_2m_2}^\psi|^2$ and ${}_{2}S_2^{\psi} = \sum_{m_1}{}_{\pm2}J_{lmlml_1m_1l_1m_1}^\psi$. 

The source of the three terms in \eq{cllensedeelss} is easy to understand. The first term corresponds to the unlensed $E$-mode spectrum. The last term comes from the combination of the $a_{lm}^X$ term with the term involving 
$J_{lml_1m_1l_2m_2l_3m_3}^\psi$. The second term arises due to the combination of the $_{\pm 2} I_{lml_1m_1l_2m_2}$ term with its own complex conjugate. Here, the identity $_{-2} I^{\psi} = (-1)^L {}_{2} I^{\psi}$ has been used. It is also easily verified that there are no other contributions at this order to the lensed power spectra. 

Next we consider the case of lensing due to gravitational waves alone ($\psi = 0$). Again, consider the computation of $ C_l^{\tilde{EE}} = (\langle {}_{+} a_{lm}^{\tilde{X}} {}_{+} a_{lm}^{*\tilde{X}} \rangle + \langle {}_{+} a_{lm}^{\tilde{X}} {}_{-} a_{lm}^{*\tilde{X}} \rangle)/2$, we arrive at the following expression for the lensed $E$-mode spectrum,
\begin{eqnarray}
\tilde{C}_{l}^{EE}&=& C_{l}^{EE}+\frac{1}{2}\sum_{l_1l_2}C_{l_1}^
{\Omega} {}_{2} S_1^{\Omega}\left[(C_{l_2}^{EE}+C_{l_2}^{BB})-(-1)^L
(C_{l_2}^{EE}-C_{l_2}^{BB})\right]\nonumber\\
&+& C_l^{EE} \sum_{l_1} C_{l_1}^{\Omega}{}_{2}S_2^{\Omega}     
\label{cllensedeegw}
\end{eqnarray} 
with ${}_{2}S_1^\Omega = \sum_{m_1m_2}|{}_{\pm 2}I_{lml_1m_1l_2m_2}^\Omega|^2$ and ${}_{2}S_2^{\Omega} = \sum_{m_1}{}_{\pm2}J_{lmlml_1m_1l_1m_1}^\Omega$. 

Again, the origin of the three terms is easily understood. The first and last terms are analogous to the corresponding terms for the case of lensing by scalars. It is in the second term that we encounter a crucial relative sign flip between two contributory terms, as compared to the LSS case. This sign flip arises from the fact that, for the case of tensor perturbations, the relevant identity is $_{\mp 2} I^{\Omega} = (-1)^{L+1} _{\pm 2} I^{\Omega}$, \eq{ilmplminus}, which is different from the corresponding relationship between $_{\mp 2} I^{\psi}$ and $_{\pm 2} I^{\psi}$, by a minus sign. Combining both, \eq{cllensedeelss} and \eq{cllensedeegw} leads us to the first equation in \eq{correctedcls}, with two additional substitutions: ${}_{2}S^{X}_1 = ({}_{2}F^{X}_{ll_1l_2})^2/(2 l + 1)$ and   \newline ${}_{2}S^{X}_2 =  -(l(l+1) - 4)l_1 (l_1 + 1) (2 l_1 + 1)/8 \pi$, $X \in \{\psi, \Omega\}$. The explicit expressions for ${}_{2}F^{X}_{ll_1l_2}$ are derived in Appendix~\ref{app:polharspace2}.

The case of the lensed $BB$ spectrum is similar to the $EE$ spectrum discussed above.
Analogously, in the case of the $TE$ cross-correlation spectrum, the identity $_{\mp 2} I^{\Omega} = (-1)^{L+1} _{\pm 2} I^{\Omega}$, leads to a relative negative sign between the terms coming from scalar and tensor perturbations respectively.
%%%%%%%%%%%%%%%%%%%%%%%%%%%%%%%%%%%%%%%%%%%%%%%%%%%%%%%%%%

%%%%%%%%%%%%%%%%%%%%%%%%%%%%%%%%%%%%%%%%%%%%%%%%%%%%%%%%%%
\section{Real space correlation function method}
\label{correlation-function-method-full-sky}
In the flat sky approximation, the sky is treated as a flat plane perpendicular to the direction $\bm{\hat{n}}$. In this limit, the spherical harmonic functions $Y_{lm}$ reduce to the well known Fourier modes,
\begin{equation}
 Y_{lm} (\mathbf{\hat{n}}) \to \frac{1}{2\pi} \, \exp(i\mathbf{l\cdot x}) \,.
\end{equation} 
We can thus write the expansion of the temperature fluctuation as follows,
\begin{equation}
 T(\mathbf{x}) = \frac{1}{2\pi} \int d^2 \mathbf{l} \, T (\mathbf{l}) \ e^{i \mathbf{l}\cdot \mathbf{x}} \,.
 \label{mlmxrelation}
\end{equation} 
Note that unlike in the case of the harmonic space method where the multipole $l$ was a discrete variable, in this case the multipole $\mathbf{l}$ is a two-dimensional wave-vector having Cartesian components  $(l \cos \phi, l \sin \phi)$ in the flat Fourier plane. Expectedly, the flat-sky analogue of the angular power spectrum is defined as follows,
\begin{equation}
\langle T (\mathbf{l})T^* (\mathbf{l'}) \rangle = C_l^{TT}\delta(\mathbf{l} -\mathbf{l'} ) \,.
\end{equation}
%The lensed temperature fluctuation at location $\mathbf{x}$ is equal to the unlensed temperature fluctuation at the location $\mathbf{x} + \bm{\alpha}(\mathbf{x})$. 
Consider first the case of lensing by scalar density perturbations, for which $\bm{\alpha} = \nabla \psi$, where $\psi$ is the lensing potential introduced earlier in \eq{psidef}.  The lensed temperature correlation function is given by the following expression,
\begin{equation}
\widetilde{\xi}(r) = \langle \widetilde{T}(\mathbf{x+\bm{\alpha}}) \widetilde{T}(\mathbf{x'+\bm{\alpha'}}) \rangle \,,
\end{equation}
where $r=|\mathbf{r}| = |\mathbf{x} - \mathbf{x'}|$ and $\widetilde{T}$ denotes the lensed temperature anisotropies on the flat sky. The ensemble average $\langle\cdots \rangle$ in the above equation is to be interpreted as an average over many realizations of the CMB sky. Since the CMB sky is assumed to be isotropic, this ensemble average could even be considered to be  over the non-overlapping patches of the CMB sky. 

The lensed correlation function can now be expressed as a function of the unlensed angular power spectrum of the temperature anisotropies $C_l^{TT}$ and the deflection field $\bm{\alpha}$ as follows,
\begin{equation}\label{lensed_corr1}
 \widetilde{\xi}(r) = \int \frac{d^2 \mathbf{l}}{(2 \pi)^2} C_l^{TT}e^{-i \mathbf{l} \cdot \mathbf{r}} \langle e^{i \mathbf{l} \cdot (\bm{\alpha'} - \bm{\alpha})} \rangle \,,
\end{equation}
where $\mathbf{l} \cdot (\bm{\alpha} - \bm{\alpha'}) = \mathbf{l} \cdot (\bm{\alpha}(\mathbf{x}) -\bm{\alpha}(\mathbf{x + r}))$.
In arriving at the above equation, another assumption which has been made is that the unlensed CMB temperature anisotropies and the projected lensing potential $\psi$ are not correlated. This allows for the averaging operation to be carried out separately on the temperature anisotropy field and the function of the deflection field.

Since the projected lensing potential is a Gaussian random field with zero mean, it is expected that the deflection field $\bm{\alpha}=\nabla \psi$ is described by the same statistics. 

For a Gaussian random variable $y$ with zero mean it can be shown that the following identity holds true,
\begin{equation}
\langle e^{iy} \rangle = \exp\Bigg[-\frac{\langle y^2 \rangle}{2} \Bigg] \,.
\end{equation}

Using the above mathematical identity, the ensemble average in \eq{lensed_corr1} can be expressed in the following form,
\begin{equation}
\langle e^{i \mathbf{l} \cdot (\bm{\alpha'} - \bm{\alpha})} \rangle = \exp \left( -\frac{1}{2} \left\langle [ \mathbf{l} \cdot (\bm{\alpha'} - \bm{\alpha})]^2 \right\rangle\right) \,.
\end{equation} 

The above equation is not in a form which allows for its numerical evaluation. To reduce the above equation to a useful form, the correlation matrix of the deflection field needs to be evaluated,
\begin{eqnarray}
 A_{ij}(\mathbf{r}) &=& \langle \alpha_i (\mathbf{x})\alpha_j (\mathbf{x+r})\rangle \nn
 &=& \langle \nabla_i \psi(\mathbf{x}) \nabla_j \psi(\mathbf{x+r}) \rangle = \int \frac{d^2\mathbf{l}}{(2\pi)^2} l_il_j C_{l}^\psi e^{i\mathbf{r\cdot l}} \label{deflection_corr_lss}
\end{eqnarray} 
where the convolution theorem for Fourier transforms has been used and $i, j = 1, 2$ denote the components of the deflection $\bm{\alpha}$. The correlation matrix of deflections $A_{ij}$ can decomposed to consist of a diagonal component and a off-diagonal component which allows for it to most generally written in the following form, 
\begin{equation}\label{gen_deflection_corr}
  A_{ij}(\mathbf{r}) = \frac{1}{2} A_0(r)\delta_{ij} - A_2(r)\left[\hat{\mathbf{r}}_i\hat{\mathbf{r}}_j - \frac{1}{2} \delta_{ij} \right] \,,
\end{equation} 
where the trace of the matrix is related to the function $A_0(r)$ and the traceless part is related to the function $A_2(r)$. Both these functions depend only on the magnitude $r = |\mathbf{r}|$ due to the assumption  of  homogeneity and isotropy.
The explicit function forms of the two functions $A_0(r)$ and $A_2(r)$ can be evaluated by equating \eq{gen_deflection_corr} \&  \eq{deflection_corr_lss} and taking the trace of $A_{ij}$ and by contracting $A_{ij}$ with $\mathbf{\hat{r}}_i \mathbf{\hat{r}}_j$ respectively \cite{durrer, lewis} to give,
\begin{equation}
 A_0(r) =  \int_0^\infty \frac{dl \ l^3}{2\pi}C_{l}^\psi J_0(rl) \,,
 \label{a0lss}
\end{equation} and
\begin{equation}
 A_2(r) = \int_0^\infty \frac{dl \ l^3}{2\pi} C_{l}^\psi J_2(rl) \,.
 \label{a2lss}
\end{equation}
Using \eq{a0lss} \& \eq{a2lss}, the expression for the lensed correlation function is given by the following expression,
\begin{equation}
\widetilde{\xi}(r) =  \int \frac{d^2 \mathbf{l}}{(2 \pi)^2} C_l^{TT}e^{-i l r \cos \phi} \exp \left[-\frac{l^2}{2}[A_0(0) - A_0(r) + A_2(r) \cos 2\phi]\right]   \,,
\label{lensedcorr}
\end{equation} 
and the lensed CMB angular power spectrum is given by the following expression,
\begin{equation}
 \widetilde{C}_l^{TT} =  \frac{1}{4\pi}  \int_0^\infty r dr J_0(rl) \widetilde{\xi}(r) \,,
 \label{lensedcl}
\end{equation} 
where $J_0$ is a Bessel function.

This completes the discussion on the evaluation of the lensed CMB angular power spectra for the case of lensing by LSS. Note that in this method it is clear that only the correlation matrix of deflection $A_{ij}$ encodes the source of lensing. For the case of lensing by GW, the evaluation of the correlation matrix $A_{ij}$ needs to be revisited while all other steps involved in arriving at the lensed CMB spectrum are identical to what has allready been discussed.

Consider the deflection field $\bm{\alpha}$  which is now defined in terms of the tensor lensing potential, $\Omega$, as $\alpha_i = \nabla_i \times \Omega = \epsilon_{ij} \nabla_j \Omega$. In complete analogy with \eq{deflection_corr_lss},  the correlation matrix of deflection arising due to lensing by GW, $A_{ij}$ can be expressed as follows,
\begin{eqnarray}
A_{ij}(\mathbf{r}) &=& \langle \alpha_i (\mathbf{x})\alpha_j (\mathbf{x+r})\rangle \nn
 &=& \langle \left( \nabla_i \times \Omega(\mathbf{x}) \right) \left(\nabla_j \times \Omega(\mathbf{x+r}) \right) \rangle =  \langle \epsilon_{ik}\nabla_k\Omega(\mathbf{x}) \epsilon_{jl}\nabla_l \Omega(\mathbf{x+r}) \rangle \nn
 &=& \int \frac{d^2\mathbf{l}}{(2\pi)^2} \epsilon_{ik}\epsilon_{jl}l_k l_l C_{l}^{\Omega \Omega} e^{i\mathbf{r\cdot l}} \label{deflection_corr_gw}
 \end{eqnarray}
%The curl deflection power spectrum $C_l^{\Omega}$ encodes all the information of the Gaussian statistics of the tensor lensing potential $\Omega$. The connection between the curl deflection power spectrum and the tensor power spectrum will be discussed in \sec{deflection_spectrum_from_power spectrum}.

The explicit function forms of the two functions $A_0(r)$ and $A_2(r)$ can be evaluated by equating \eq{gen_deflection_corr} \&  \eq{deflection_corr_gw} and taking the trace of $A_{ij}$ and by contracting $A_{ij}$ with $\mathbf{\hat{r}}_i \mathbf{\hat{r}}_j$ respectively.

Taking the trace of $A_{ij}$ results in the following expression,
\begin{equation}
 A_0(r) = A_{ii} = \epsilon_{ik} \epsilon_{il}\int \frac{d^2\mathbf{l}}{(2\pi)^2}\, l_kl_l\, C_l^{\Omega} \, e^{i\mathbf{r\cdot l}} 
 = \int \frac{d^2\mathbf{l}}{(2\pi)^2}\, l^2\, C_l^{\Omega}\, e^{i\mathbf{r\cdot l}} \,,
 \label{a0gw}
\end{equation} 
in which we have used $\epsilon_{ij} \epsilon_{ib} = \delta_{jb}$. Note that $A_0(r)$ derived in the case of lensing by GW has the same form as that for the case of lensing by LSS, except that $C_l^{\psi}$ is replaced by $C_l^{\Omega}$.

To find $A_2(r)$, the correlation matrix $A_{ij}$ is contracted with  $\hat{\mathbf{r}}_i\hat{\mathbf{r}}_j$ to get,
\begin{eqnarray}
 A_{ij} \hat{\mathbf{r}}_i\hat{\mathbf{r}}_j &=&  \epsilon_{ik} \epsilon_{jl}\int \frac{d^2\mathbf{l}}{(2\pi)^2}\, l_kl_l\,\hat{\mathbf{r}}_i\hat{\mathbf{r}}_j  \, C_l^{\Omega}\, e^{i\mathbf{r\cdot l}} \nonumber\\
 &=& \int  \frac{d^2\mathbf{l}}{(2\pi)^2} (\mathbf{\hat{r}}\times\mathbf{l})^2  C_l^{\Omega}\, e^{i\mathbf{r\cdot l}} \,,
\end{eqnarray}
where we have the analogue of the cross product in two dimensions, $\hat{\mathbf{r}} \times \mathbf{l}$ being given by $\epsilon_{ij}\mathbf{\hat{r}}_i l_j$. This can be replaced by $l  \sin \phi$ where $\phi$ is the angle between $\mathbf{l}$ and $\mathbf{\hat{r}}$. 
%The above form of the expression provides a natural intuitive extension of the (small-angle) result in Ref. \cite{cooray} for the arbitrary deflection angle case, in which we again see that a dot product between two terms in case of lensing by scalars is replaced by a cross product in the case of lensing by tensors. 
Further simplification of the above equation results in the following result,
\begin{eqnarray}
A_{ij} \mathbf{\hat{r}}_i  \mathbf{\hat{r}}_j &=& \frac{1}{2} (A_0 - A_2(r)) \label{local1}\\
&=& \int_0^\infty  \frac{dl}{(2\pi)^2}l^3 C_l^{\Omega}\int_0^{2\pi} \sin^2\phi \, e^{ilr\cos\phi} d\phi \,, \nonumber \\
 &=& \int_{0}^{\infty} \frac{dl}{(2\pi)^2} l^3  C_{l}^\Omega\int_0^{2\pi}d\phi\,\frac{[1 - \cos(2\phi)]}{2} \,e^{ilr\cos\phi} \nonumber \\
 &=& \frac{1}{2} \int_0^\infty \frac{dl \ l^3}{2\pi} C_{l}^\Omega (J_0(rl) + J_2(rl)) \label{local2}\,.
\end{eqnarray} 
Comparing \eq{local1} with \eq{local2} reveals the functional form of the function $A_2(r)$ which is the following,
\begin{equation}
 A_2(r) = - \int_0^\infty \frac{dl}{2\pi} \, l^3  C_l^{\Omega} \, J_2({rl}) \,.
 \label{a2gw}
\end{equation} 

Now that the functional forms of $A_0$ and $A_2$ are known, it is possible to evaluate the lensed CMB angular power spectrum due to lensing by GW's.

Comparing \eq{a0lss} \& \eq{a2lss} with \eq{a0gw} \& \eq{a2gw} respectively, reveals the following points,
\begin{itemize}
\item The function $A_0$ has identical form for both lensing due to LSS as well as lensing due to GW. 
\item The functional form of $A_2$ for lensing due to GW has an additional negative sign relative to $A_2$ due to lensing by LSS.
\item The evaluation of the functions $A_0$ and $A_2$ for the case of lensing by GW requires the tensor lensing potential power spectrum $C_l^{\Omega}$ whereas the evaluation of the functions $A_0$ and $A_2$ for the case of lensing by LSS requires the  lensing potential power spectrum $C_l^{\psi\psi}$.
\end{itemize}

The above list of observation also summarises the similarity between the lensing kernels arising due to lensing by GW and LSS. The function $A_2$ only differing by a relative sign difference for the two cases of lensing by LSS and GW is very reminiscent of similar observations made in the harmonic space lensing kernels.

\subsection{Evaluation of the lensed CMB polarization power spectra}
To evaluate the effect of lensing on the polarization power spectra requires the evaluation of the lensed correlation function defined over the polarization field. The evaluation of the correlation matrix of deflection remains identical to what has already been discussed for the case of lensing of CMB temperature anisotropies. The details of the calculation for the case of polarization are not provided, however the final results are stated for completeness.

To arrive at the lensing modifications to the polarization power spectra $C_l^{EE}$ \& $C_l^{BB}$ requires the evaluation of the following correlation functions,
\begin{eqnarray}
\widetilde \xi_+(r) 
&=&\frac{1}{2\pi} \int l \, dl \,\left[ C_l^{EE} +  C_l^{BB}\right] \, e^{-(l^2/2) (A_0(0) - A_0(r))}\nonumber\\
 &\times& \left( I_0(l^2 A_2(r)/2)\ J_0(rl) + 2 \sum_{n=1}^\infty I_n(l^2A_2(r)/2) \, J_{2n}(rl)\right) \,, \\
 \widetilde \xi_-(r) & = & \int\frac{l\,dl}{2\pi}  \left[ C_l^{EE} -  C_l^{BB}\right] \, e^{-l^2 (A_0(0) - A_0(r))/2 }\nonumber\\
 \qquad &\times& \Big( \left[ 2 I_0''\,  (l^2 A_2(r)/2) - I_0 (l^2 A_2(r)/2) \right] \, J_0(lr)\nonumber\\
\qquad &+& 2 \sum_{n=1}^\infty \left[ 2 I_n''\,  (l^2 A_2(r)/2) - I_n (l^2 A_2(r)/2) \right] \, J_{2n}(lr)\Big) \,,
\end{eqnarray}
and to calculate the lensing modifications to the temperature-polarization cross power spectrum $C_l^{TE}$ requires the evaluation of the following correlation function,
\begin{eqnarray}
\widetilde \xi_\times(r) & = & \int\frac{l\,dl}{2\pi} C_l^{TE}\,  e^{-l^2 (A_0(0) - A_0(r))/2 } \nn
\qquad &\times& \left( I_0' (l^2 A_2(r)/2) \, J_0(lr) + 2 \sum_{n=1}^\infty (-1)^n I_n' (l^2 A_2(r)/2)\, J_{2n}(lr)\right) \,.
\end{eqnarray}

In the above expressions, the function $I_n$ is the modified Bessel function, and the prime denotes differentiation with respect to the argument. While evaluating lensing modifications to the CMB spectra, in the case of lensing by LSS the form of $A_0$ and $A_2$ defined in Eqs. (\ref{a0lss}) and (\ref{a2lss}) are substituted in the above equations whereas in the case of lensing by GW  the form of $A_0$ and $A_2$ defined in Eqs. (\ref{a0gw}) and (\ref{a2gw}) are substituted.

Finally to arrive at the lensed CMB angular power spectra the following relations are to be used,
\begin{eqnarray}
\tilde{C}_l^{EE}+\tilde{C}_l^{BB}=2\pi \int dr J_0(rl) \tilde{\xi}_+(r) \,, \\
\tilde{C}_l^{EE}-\tilde{C}_l^{BB}=2\pi \int dr J_4(rl) \tilde{\xi}_-(r) \,, \\
C_l^{TE}=2\pi \int dr J_2(rl) \tilde{\xi}_\times(r) \,.
\end{eqnarray}
%%%%%%%%%%%%%%%%%%%%%%%%%%%%%%%%%%%%%%%%%%%%%%%%%%%%%%%%%%
%\input{references}

%%%%%%%%%%%%%%%%%%%%%%%%%%%%%%%%%%%%%%%%%%%%%%%%%%%%%%%%%%
\end{document}